\documentclass{aa}
\usepackage{txfonts}
\usepackage{natbib}
\usepackage{supertabular}
\usepackage{lscape}
\usepackage{indentfirst}
\usepackage{aalongtable}
\usepackage{graphicx}

\bibpunct{(}{)}{;}{a}{}{,}

\begin{document}

\title{Multi-wavelength properties of \textit{Spitzer} selected starbursts at z\,$\sim$\,2 \thanks{color figures and the Appendix are only available in the electronic form via http://www.edpsciences.org}}

\author{N. Fiolet\inst{1,2}
  \and A. Omont\inst{1,2}
  \and M. Polletta\inst{1,2,3}
  \and F. Owen\inst{4}
  \and S. Berta\inst{5}
  \and D. Shupe\inst{6}
  \and B. Siana\inst{7} 
  \and C. Lonsdale\inst{8}
  \and V. Strazzullo\inst{4}
  \and M. Pannella\inst{4}
  \and A.~J. Baker\inst{9}
  \and A. Beelen\inst{10}
  \and A. Biggs\inst{11,12}
  \and C. De Breuck\inst{12}
  \and D. Farrah\inst{13}
  \and R. Ivison\inst{11, 14}
  \and G. Lagache\inst{10}
  \and D. Lutz\inst{5}
  \and L.~J. Tacconi\inst{5}
  \and R. Zylka\inst{15}
   }


\institute{UPMC Univ Paris 06, UMR7095, Institut d'Astrophysique de Paris, F-75014, Paris, France 
\and CNRS, UMR7095, Institut d'Astrophysique de Paris, F-75014, Paris, France 
\and INAF-IASF Milano, via E. Bassini 15, 20133 Milan, Italy 
\and National Radio Astronomy Observatory, P. O. Box 0, Socorro, NM 87801, USA 
\and Max-Planck Institut f\"ur extraterrestrische Physik, Postfach 1312, 85741 Garching, Germany 
\and Herschel Science Center, California Institute of Technology, 100-22, Pasadena, CA 91125, USA 
\and Astronomy Department, California Institute of Technology, MC 105-24, 1200 East California Boulevard, Pasadena, CA 91125, USA 
\and North American ALMA Science Center, NRAO, Charlottesville, USA 
\and Department of Physics and Astronomy, Rutgers, the State University of New Jersey, 136 Frelinghuysen Road, Piscataway, NJ 08854, USA 
\and Institut d'Astrophysique Spatiale, Universit\'e de Paris XI, 91405 Orsay Cedex, France 
\and UK Astronomy Technology Centre, Royal Observatory, Blackford Hill, Edinburgh EH9 3HJ 
\and European Southern Observatory, Karl-Schwarzschild Strasse, 85748 Garching bei M\"unchen, Germany 
\and Department of Physics \& Astronomy, University of Sussex, Falmer, Brighton, BN1 9RH, UK
\and Institute for Astronomy, University of Edinburgh, Blackford Hill, Edinburgh EH9 3HJ 
\and Institut de Radioastronomie Millim\'etrique, 300 rue de la Piscine, 38406 St. Martin d'H\`eres, France}

\abstract{Wide-field \textit{Spitzer} surveys allow identification of 
thousands of potentially high-$z$ submillimeter galaxies (SMGs) through their 
bright 24\,$\mu$m emission and their mid-IR colors.}{We want to determine the 
average properties of such $z\sim$2 \textit{Spitzer}-selected SMGs by 
combining millimeter, radio, and infrared photometry for a representative 
IR-flux ($\lambda_{\rm rest}\sim 8\,\mu$m) limited sample of SMG 
candidates.}{A complete sample of 33 sources believed to be starbursts 
(``5.8\,$\mu$m-peakers'') was selected in the (0.5\,deg$^2$) J1046+56 field  with 
selection criteria $F_{\rm 24\,\mu m}$\,\textgreater\,400\,$\mu$Jy, the 
presence of a redshifted stellar emission peak at 5.8\,$\mu$m, and 
$r^\prime_{\rm Vega}$\,\textgreater\,23. The field, part of the SWIRE Lockman Hole field, benefits from very deep VLA/GMRT 20\,cm, 50\,cm, and 90\,cm 
radio data (all 33 sources are detected at 50\,cm), and deep 160\,$\mu$m and 
70\,$\mu$m \textit{Spitzer} data. The 33 sources, with photometric redshifts 
$\sim1.5\,-\,2.5$, were observed at 1.2\,mm with IRAM-30m/MAMBO to an 
rms $\sim$0.7\,-\,0.8\,mJy in most cases. Their millimeter, radio, 7-band 
\textit{Spitzer}, and near-IR properties were jointly analyzed.}{ The 
entire sample of 33 sources has an average 1.2\,mm flux density of $1.56 
\pm 0.22$\,mJy and a median of 1.61\,mJy, so the majority of the sources 
can be considered SMGs. Four sources have confirmed 4\,$\sigma$ 
detections, and nine were tentatively detected at the 3\,$\sigma$ level. 
Because of its 24\,$\mu$m selection, our sample shows systematically lower 
$F_{\rm 1.2\,mm}/F_{\rm 24\,\mu m}$ flux ratios than classical SMGs, probably because of enhanced PAH emission. A median FIR 
SED was built by stacking images at the positions of 
21 sources in the region of deepest \textit{Spitzer} coverage. Its parameters are $T_{\rm dust} = 37 \pm 8$\,K, $L_{\rm FIR} = 
2.5 \times 10^{12}\,L_{\odot}$, and SFR\,=\,450\,$M_{\odot}$\,yr$^{-1}$. The FIR-radio correlation provides another estimate of $L_{\rm FIR}$ for each 
source, with an average value of  $4.1 \times 10^{12}\,L_{\odot}$; however, 
this value may be overestimated because of some AGN contribution.
Most of our targets are also luminous star-forming $BzK$ galaxies which 
constitute a significant fraction of weak SMGs at $1.7 \lesssim z \lesssim 
2.3.$}{\textit{Spitzer} 24\,$\mu$m-selected starbursts and AGN-dominated 
ULIRGs can be reliably distinguished using IRAC-24\,$\mu$m SEDs. Such 
``5.8\,$\mu$m-peakers'' with $F_{\rm 24\mu m}$\,\textgreater\,400\,$\mu$Jy 
have $L_{\rm FIR}\,\gtrsim 10^{12}\,L_{\odot}$. They are thus $z \sim 2$ 
ULIRGs, and the majority may be considered SMGs. However, they have 
systematically lower 1.2\,mm/24\,$\mu$m flux density ratios than classical 
SMGs, warmer dust, comparable or lower IR/mm luminosities, and higher stellar 
masses. About 2000\,$-$\,3000 ``5.8\,$\mu$m-peakers'' may be easily 
identifiable within SWIRE catalogues over 49\,deg$^2$.}

\maketitle

\section{Introduction}

Ultra-Luminous InfraRed Galaxies (ULIRGs, with $L_{\rm FIR} \gtrsim 
10^{12}\,L_{\odot}$) are the most powerful class of star-forming galaxies. For 
25 years, these prominent sources and their intense starbursts have been the 
target of many comprehensive studies, both locally 
\citep[e.g., ][]{Sand96,Lons06,Veil09} and at high redshift 
\citep[e.g., ][]{Blai04,Solo05}.  While local ULIRGs are relatively rare, 
submm/mm surveys with large bolometer arrays such as JCMT/SCUBA 
[James Clerk Maxwell Telescope/Submillimetre Common User Bolometer Array \citep[]{Holl99}], 
 APEX/LABOCA [Atacama Pathfinder Experiment/Large Apex Bolometer Camera \citep[]{Siri09}] 
 or IRAM/MAMBO [Institut de Radioastronomie Millim\`etrique/Max-Planck Bolometer Array \citep[]{Krey98}]
 have shown that the como\-ving density of submillimetre galaxies (SMGs), which 
represent a significant class of high-redshift ($z\sim1-4$) ULIRGs,
 is about a thousand times greater than that of ULIRGs in the local Universe 
\citep[e.g.,][]{lefl05,Chap05}. 
They represent a major phase of star formation at early epochs and are also 
characterized by high stellar masses \citep[e.g.,][]{Bory05}. 
They are thus ideal candidates to be the precursors of local massive 
elliptical galaxies \citep[e.g.,][ hereafter Lo09, and references 
therein]{Blai02,Dye08,Lons08}.  Nearly all of the enormous UV energy produced 
by their massive young stars is absorbed 
by interstellar dust and re-emitted at far-infrared wavelengths, with their far-infrared 
luminosity ($L_{\rm FIR}$) able to reach $10^{13}\,L_{\sun}$. 
However, despite the considerable efforts invested in mm/submm surveys, the 
total number of known SMGs remains limited to several hundred, and current 
observational capabilities are still somewhat marginal at many wavelengths.
 We thus still lack comprehensive studies of SMGs and their various subclasses 
at all wavelengths and redshifts and in various environments. 
Even their star formation rates (SFRs) remain uncertain in most cases because 
of a lack of direct observations at the FIR/submm wavelengths of their maximum 
emission.
 The identification of large samples of SMGs is important for investigating 
the properties of these galaxies (SFR, luminosity, spectral energy 
distribution [SED],
 stellar mass, AGN content, spatial structure, radio and X-ray parameters, 
clustering, etc.) on a statistical basis, as a function of their various 
subclasses, redshift, and environment. This is the main goal of the wide-field submm 
surveys planned with SCUBA2 and \textit{Herschel}.

Although \textit{Spitzer} generally lacks the sensitivity to detect SMGs in 
the far-IR, its very good sensitivity in the mid-IR allows the efficient 
detection of a significant fraction of SMGs in the very large area observed 
by its wide-field surveys, and in particular the $\sim49\,\rm{deg}^2$ {\it Spitzer} 
Wide-area Infrared Extragalactic (SWIRE) survey \citep{Lons03}. 
From an analysis of a sample of $\sim 100$ SMGs observed with 
\textit{Spitzer}, Lo09 have estimated that SWIRE has detected
 more than 180 SMGs with $F_{\rm 1.2\,mm} > 2.5\,$mJy per square degree at 
24\,$\mu$m and in several IRAC bands from 3.6 to 8.0\,$\mu$m. However, the 
identification of SMGs among SWIRE sources is not straightforward, since it 
requires inferring FIR emission from mid-IR photometry in objects with various 
SEDs, especially as regards AGN versus starburst emission, and various 
redshifts.

We have therefore undertaken a systematic study of the 1.2\,mm emission from the 
best SMG candidates among \textit{Spitzer} bright 24\,$\mu$m sources, focusing 
on $z\,\sim$\,2 starburst candidates.  In \citet{Lons06} and Lo09 \citep[see 
also][]{Weed06b,Farr08}, it is shown that selecting sources with a secondary 
maximum emission in one of the intermediate IRAC bands at 4.5 or 5.8\,$\mu$m 
provides an efficient discrimination against AGN power-law SEDs. In 
particular, 24\,$\mu$m bright ``5.8\,$\mu$m-peakers'' have a high probability 
of being dominated by a strong starburst at $z\sim 2$, whose intense 7.7\,$\mu$m 
feature is redshifted into the 24\,$\mu$m band. A first 1.2\,mm MAMBO study of a 
sample of $\sim 60$ bright SWIRE sources (Lo09) has confirmed that such a 
selection yields a high detection rate at 1.2\,mm and a significant average 
1.2\,mm flux density, showing that the majority of such sources are high-$z$ 
ULIRGs, probably at $z\sim 2$. However, as described in Lo09, this sample was 
selected with the aim of trying to observe the ``5.8\,$\mu$m-peakers'' 
with the strongest mm flux over more than 10\,deg$^{2}$. This was achieved by 
deriving photometric redshifts, estimating the expected 1.2\,mm flux densities 
by fitting templates of various local starbursts and ULIRGs to the optical and 
infrared (3.6$-$24\,$\mu$m) bands, and selecting the candidates predicted to 
give the strongest mm emission. Therefore, the selection criteria of this 
sample were biased, especially toward the strongest 24\,$\mu$m sources and 
those in clean environments.  We report here the results of an analogous MAMBO 
study, but of a complete 24\,$\mu$m-flux limited sample of all SWIRE 
``5.8\,$\mu$m-peakers'' in a 0.5\,deg$^{2}$ region within the SWIRE Lockman 
Hole field, with $F_{\rm 24\,\mu m} > 400\,\mu$Jy and $r^\prime_{\rm Vega} > 
23$ 
(see Sec.\ 2 for a precise definition of ``5.8\,$\mu$m-peakers'', which of 
course depends on the actual SWIRE data and limits of sensitivity and 
accuracy). 
This region was selected because of the richness in multi-wavelength data, in 
particular the exceptionally deep radio data at 20\,cm \citep[VLA,][]{OwMo09}, 
 50\,cm (GMRT, Owen et al. in prep.), and 90\,cm \citep[VLA, ][]{Owen09}. Our 
study aims at characterizing the average multi-wavelength properties of these 
sources, their dominant emission processes (starburst or AGN), their stellar 
masses, and their star formation rates. 
We adopt a standard flat cosmology: $H_{0}$=71\,km\,s$^{-1}$\,Mpc$^{-1}$, 
$\Omega_{M}$=0.27 and $\Omega_{\Lambda}$=0.73 \citep{Sper03}. 

\section{Sample selection and ancillary data}\label{sample}

\begin{figure}[!htbp]
\resizebox{\hsize}{!}{\includegraphics{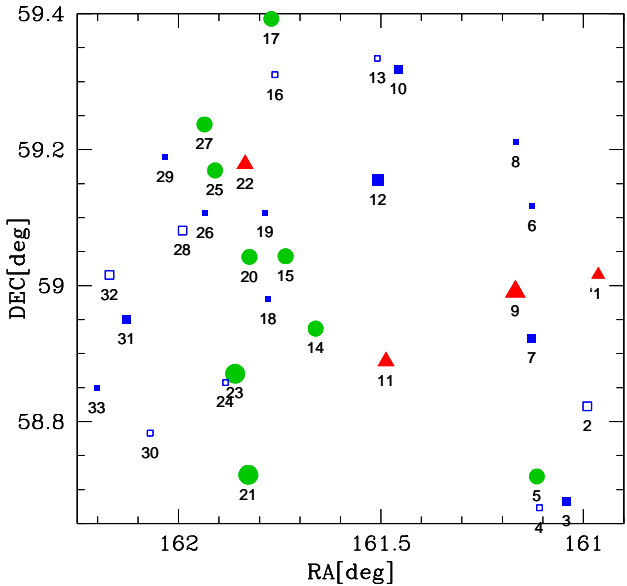}}
\caption{Positions of the 33 sources of our sample ($42^\prime \times 
42^\prime$ J1046+59 field, centred at the VLA position observed by 
\citet{OwMo09}, $\alpha_{2000}=10$h$46$m$00$s, 
$\delta_{2000}=+59\degr01\arcmin00\arcsec$). The red filled triangles are the 
$4\sigma$ detections at 1.2\,mm.
 The green filled circles are the $3\sigma$ tentative detections at 1.2\,mm. 
The blue squares are sources with a signal at 1.2\,mm lower than $3\sigma$. 
The sizes of symbols are proportional to 1.2\,mm strengths. The open 
symbols show no detection at 20\,cm 
 \citep{OwMo09}.}
\label{map.eps}
\end{figure}

We selected all \textit{Spitzer}/SWIRE ``5.8\,$\mu$m-peakers'' with $F_{\rm 
24\,\mu m} > 400\,\mu$Jy in the $42^\prime \times 42^\prime$ (0.49\,deg$^2$) 
J1046+59 field in the SWIRE Lockman
Hole, centered at $\alpha_{2000}=10$h$46$m$00$s,
$\delta_{2000}=+59\degr01\arcmin00\arcsec$ (Fig.\ref{map.eps}). A source is 
considered to be a ``5.8\,$\mu$m-peaker'' if it satisfies the following 
conditions: $F_{\rm 3.6\,\mu m} < F_{\rm 4.5\,\mu m} < F_{\rm 5.8\,\mu 
m} > F_{\rm 8.0\,\mu m}$, without consideration of uncertainties.  13 sources 
have no detection in the 8.0\,$\mu$m band.  We assume that these sources are 
also ``5.8\,$\mu$m-peakers'' because their fluxes at 5.8\,$\mu$m are greater 
than the detection limits at 8.0\,$\mu$m (\textless\,40\,$\mu$Jy). We also 
require that the sources are optically faint, i.e., $r^\prime_{\rm Vega} > 23$,
to remove low redshift interlopers
\citep{Lons06}. The selected sample contains 33 sources, which represent
6\% of all sources with $F_{\rm 24\,\mu m} > 400\,\mu$Jy and
$r^\prime_{\rm Vega} > 23$ in the field.

We use the SWIRE internal catalogue available at the time of definition of
the project, September 2006. Details on the SWIRE observations and data are
available in~\citet{Sura05}. The 2006 SWIRE internal catalogue has been
superseded by the current version, dated 2007.  
 Since our selection, MAMBO observations, and analysis were based on the
2006 catalogue, but there are no significant changes for our sources in the
2007 catalogue, and the numbers of sources selected in the two versions of
 the catalogue do not vary signifi\-cantly, for this work we will use the selection from 
2006 data. Based on the analysis of the sources that would have been missed
or included applying our selection criteria to the two versions of the
catalogues, we find that the original sample selected from the 2006 SWIRE
catalogue remains representative of a sample strictly meeting our selection
criteria.

We have reported in Table~\ref{new} the fluxes from the 2006 SWIRE catalogue.
The optical magnitudes have been obtained with the MOSAIC camera on the 4$-$meter Mayall Telescope at Kitt Peak National Observatory \citep[e. g., ][]{Mull98}.
 However, for a few sources, the optical data differ slightly from those available in the
SWIRE catalogue because a measurement at each source position was performed
 for all non-detections in the cata\-logue.   
 The revised optical magnitudes
are listed in Table~\ref{new}. The imaging in the central 0.3\,deg$^{2}$ is much deeper (by $\sim0.7$\,mag). Exact depths are listed in \citet{Poll06}. 
In addition to the SWIRE data, we obtained near-infrared (NIR) data from two sources: 1) from Palomar/WIRC \citep{Wils03}; 2) from
UKIRT/WFCAM \citep{Henr00} in the $J$ ($\lambda$\,=\,1.25\,$\mu$m), $H$ ($\lambda$\,=\,1.63\,$\mu$m), and $K$
($\lambda$\,=\,2.20\,$\mu$m) bands as part of a NIR survey of the field. A
description of these data and their reduction will be published in
Strazzullo et al., in prep. The NIR data are reported in
Table~\ref{NIR}. 23 sources are detected in all three bands. 24 sources are detected at least in the $K$-band. The WFCAM $K$ band data are public data from UKIDSS \citep{Lawr07,Warr07}.
 Pipeline processing and the science archive are described by \citet{Hamb08} and Irwin et al.~(in prep.).

Table~\ref{samples} compares the selection criteria for our sample to those for si\-milar
samples of bright 24\,$\mu$m sources. The sample of Lo09 is based on the
same criteria, but is biased toward sources brighter at 24\,$\mu$m, with
819\,$\mu$Jy on average vs. 566\,$\mu$Jy for the present
sample. The sample of \citet{Farr08} is similar but aimed at
``4.5\,$\mu$m-peakers''; the sample of \citet{Huan08} and
\citet{Youn09}
is based on different IRAC criteria, but they indeed select
almost exclusively ``5.8\,$\mu$m-peakers'' (Sec. 5.1). On the other hand, the
selection criteria of \citet{Magl07} and \citet{Yan05}, which do not use the
IRAC flux densities, do not discriminate against AGN and yield a large
proportion of AGN.

Compared to the twin starburst sample of Lo09, the present sample is
complete down to a 24\,$\mu$m flux density of 400\,$\mu$Jy. It thus includes
weaker 24\,$\mu$m sources on average, but it should be free from the 
 selection biases present in the Lo09 sample that resulted from the
effort to maximize the number of detections at 1.2\,mm. In addition, our sample
benefits from very deep radio data at 1400, 610, and 324\,MHz (Sec 4.3). 
The positions of the sources in the field are shown in Figure~\ref{map.eps}. The radio flux densities are listed in Table~\ref{res}.

\section{MAMBO observations and results}

Observations were made during the winter 2006/2007 MAMBO observing pool 
between December 2006 and March 2007 at the IRAM 30\,m telescope, 
located at Pico Veleta, Spain, using the 117-element version of the MAMBO 
array \citep{Krey98} operating at 1.2\,mm (250\,GHz). 
We used a standard ``on-off'' photometry observing mode with a secon\-dary 
mirror wobbling at a frequency of 0.5\,Hz between the source 
and a blank sky position offset in azimuth by $\sim\,35\arcsec$. Periodically, 
the telescope was nodded so that the sky position lay on the other side of the source position. 
 Pointing and focus were updated regularly on standard sources. 
Nearly every hour, the atmospheric opacity was measured by observing the sky at six elevations. Our observations are divided in 16 or 20 ``subscans'' 
of 60 seconds each. 
In this operating mode, the integration time is $\sim$\,40\,s (20\,s on source and 20\,s on sky) per subscan.  Observations of each source were never concentrated in a single night, 
  but distributed over several nights in order to reduce the risks of systematic effects.  
 The initial aim was to observe the 32 sources (L-12 was observed in the project described by Lo09, under the name ``LH-11'') with an rms $\sim$\,0.8\,mJy,  
which corres\-ponds to $\sim$\,0.6\,h of integration for the system sensitivity $\sim$\,35-40\,mJy\,s$^{-1/2}$ in average weather conditions. 
As seen in Table~\ref{res}, this was achieved for most of the sources. However, for $\sim$\,20\% of the sources the rms was instead $\sim$\,0.9-1.1\,mJy, 
while a similar number of sources were observed somewhat longer to reach an rms $\sim$\,0.5-0.6\,mJy in order to confirm their detection. 

The data reduction is straightforward thanks to the MOPSIC software
package\footnote{Documentation by R. Zylka is available at http://www.iram.es/IRAMES/mainWiki/CookbookMopsic}, which is regularly updated on the MAMBO pool page.
 This program reduces the noise due to the sky emission if
it is sufficiently correlated between the diffe\-rent bolometers. This process is gene\-rally sufficient for the majority of
observations. However, in a few cases, some scans may present faults due, e.g.,
to lack of helium in the cryostat or problems of acquisition. These 
scans are validated or rejected after close inspection.

The results of our observations at 1.2\,mm are reported in Table \ref{res}.
 The average flux density (with equal weight) of the entire sample is
$1.56 \pm 0.22$\,mJy, very comparable to Lo09 ($1.49 \pm 0.18$\,mJy) and 
\citet{Youn09} ($1.6 \pm 0.1$\,mJy), but greater than obtained 
by \citet{Lutz05} for a {\it Spitzer}-selected sample of high-$z$ starbursts 
and (mostly) AGNs ($0.74 \pm 0.09$\,mJy).
  The median for our sample is 1.61\,mJy. This confirms that on ave\-rage 
the majority of these sources are 
SMGs~\citep[at $z \sim 2$, 1.6\,mJy cor\-responds to $\sim$4\,mJy at
850\,$\mu$m:][]{Grev04}. Because of the limited integration time, only 
four sources were solidly detected at $>$4$\,\sigma$. However, the fraction 
of sources at least tentatively detected 
 at $>$ 3$\sigma$ is 39\% (13/33 sources). It is worth stressing that the reliability of such $3\sigma$ tentative detections in careful ''on-off'' 
 MAMBO observations is much higher than those detected in a mm/submm map amongst 
 hundreds of possible resolution elements, where flux boosting is inevitable.

This on-off $3\sigma$ 'detection' rate is similar to what was obtained by Lo09 for a similar sample, 
 higher than obtained by \citet{Lutz05} for their 
{\it Spitzer}-selected sample of starbursts and (mostly) AGNs (18\%), and
lower than obtained by~\citet{Youn09} with deeper observation of a 
similar sample of {\it Spitzer}-selected $z \sim 2$ starbursts (75\%).

\section{Source properties}

\subsection{Spectral energy distributions and redshifts}

\begin{figure}[!htbp]
\resizebox{\hsize}{!}{\includegraphics{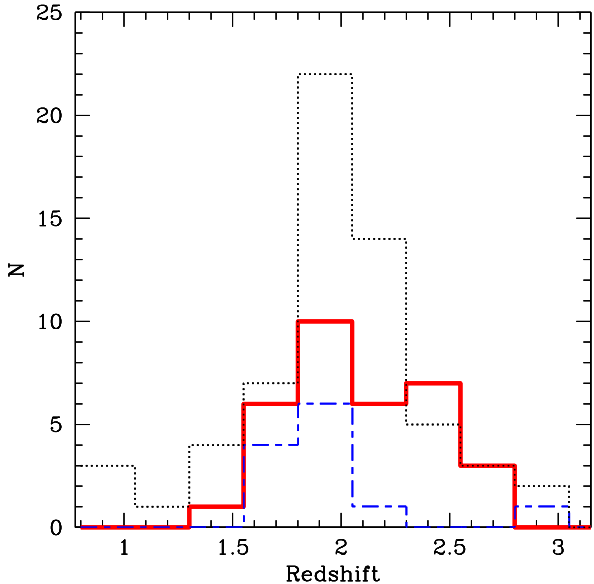}}
\caption{Histogram of redshifts for our sample (photometric, thick solid red line), the full sample from Lo09 (photometric and spectroscopic, dotted black line), and the sample from \citet{Youn09} (spectroscopic, long-short-dashed blue line).}
\label{histo.eps}
\end{figure}

In order to estimate photometric redshifts, we fit the spectral energy
distributions (SEDs), including optical and IR ($\le 24\,\mu$m) data for each
source, with a library of galaxy templates following the method described in
 Lo09 and \citet{Poll07}. The SEDs are fitted using the Hyper-$z$
code~\citep{Bolz00}, and the effects of dust extinction are taken into
account. As discussed in Lo09, such photometric redshifts are li\-mited in
accuracy and have uncertainties of $\pm$0.5.

SED fits for all the sources in the sample are shown in the Appendix (only in the electronic edition), and the
photometric redshifts are listed in Table \ref{res}. The photometric
redshift distribution of the sample is shown in Fig.~\ref{histo.eps}. All
redshifts but four are within the range $1.5 \lesssim z \lesssim 2.5$. The 
average
redshift from these SEDs is $<\,z\,> = 2.08$ (median\,=\,2.04, scatter\,=\,0.32, and semi-inter-quartile range\,=\,0.26). This result is consistent
with our selection criteria, which assume that the 7.7\,$\mu$m PAH band is
redshifted into the 24\,$\mu$m MIPS band and the 1.6\,$\mu$m stellar band into
the 5.8\,$\mu$m IRAC band. The redshift distribution of our sample is similar
to the one measured in Lo09 ($<\,z\,> = 1.97 \pm 0.05$), which is mostly based on
photometric redshifts, and the one measured
in~\citet{Youn09} ($<\,z\,> = 1.96 \pm 0.10$), which is based on spectroscopic
redshifts (Fig.~\ref{histo.eps}).
Thus, all these works select sources in a similar redshift range.
Indeed, the actual redshift distribution of our sample might be similar to that of \citet{Huan08} and \citet{Youn09} 
and concentrated within  a rather narrow redshift range around $z\sim$\,1.7-2.3 as shown by the few spectroscopic redshifts 
reported for the sample of Lo09. 

Twelve sources from our sample have redshifts from the catalogue of SWIRE photometric redshifts of \citet{Rowa08}. 
For ten of these sources, our photometric redshifts determined by the SED fitting show a good ($\pm\,10\%$) agreement with the determinations from \citet{Rowa08}.

\subsection{Comparison between 1.2\,mm and 24\,$\mu$m flux densities}

\begin{figure}[!htbp]
\resizebox{\hsize}{!}{\includegraphics{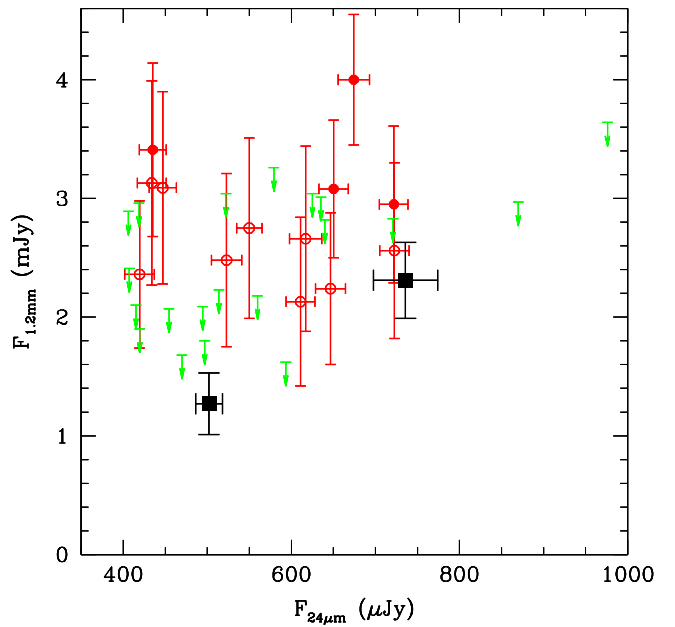}}
\caption{Observed MAMBO 1.2\,mm flux density as a function of 24\,$\mu$m 
flux density.  The black squares show the average flux densities for the first
$F_{\rm 24\,\mu m}$ quartile (8 sources) and for the 25 other sources, respectively.
 The filled (open) red circles are the sources that are $4\sigma$ detections 
($3\sigma$ tentative detections) at 1.2\,mm.  
 The green arrows are 2\,$\sigma$ upper limits for the sources that 
are not detected at 1.2\,mm.
 }
\label{1224.eps}
\end{figure}

In order to investigate whether there is a correlation between mid-IR and mm
emission, we compare the flux densities at 1.2\,mm and at 24\,$\mu$m
($F_{\rm 1.2\,mm}$ and
$F_{\rm 24\,\mu m}$) in Figure~\ref{1224.eps}.  Because of the limited
sensitivity of the 1.2\,mm data, we have stacked the data for the first
$F_{\rm 24\,\mu m}$ quartile (8 sources), and independently for the 25 other sources. These stacked values have been computed with the observed $F_{\rm 1.2\,mm}$ values. 
 Fig. \ref{1224.eps} shows that it is impossible to see whether there
is a correlation between mid-IR and mm emission with this sensitivity. However, the average
for the highest $F_{\rm 24\,\mu m}$ quartile seems $\sim 1.5-2$ times larger than the average for all the other sources.

Because of the 24\,$\mu$m selection, the  $F_{\rm 1.2\,mm}/F_{\rm 24\,\mu
m}$ ratio of our sources is relatively low compared to that submm selected
SMGs, as in the case of Lo09.
The ratio of the average $F_{\rm 1.2\,mm}$ to the average $F_{\rm 24\,\mu m}$ is $2.76 \pm 0.50$ for
the entire sample of 33 sources (and $4.94 \pm 0.51$ for 13 sources with
1.2mm S/N\,$>$\,3, see Table 3).
  As seen in Fig.~\ref{1224z1.eps},  these ratios of averages are a factor
$\sim 4$ ($\sim 2$) smaller 
than that of a sample of literature SMGs (Sec. 5.2),
 and a factor $\sim 6$ ($\sim 10$) larger than one of the (AGN dominated)
sample of bright 24\,$\mu$m sources of  \citet{Lutz05}. 
They are a factor $\sim 10$ ($\sim 5$) smaller than that of the extreme SMG
template Arp 220 at $z=2$,
but comparable to those of starburst templates (M\,82 or NGC\,6090) and
AGN-starburst composites (IRAS19254$-$7245 South).

\subsection{Radio properties and the nature of the sources}\label{radio_prop}

\begin{figure}[!htbp]
\resizebox{\hsize}{!}{\includegraphics{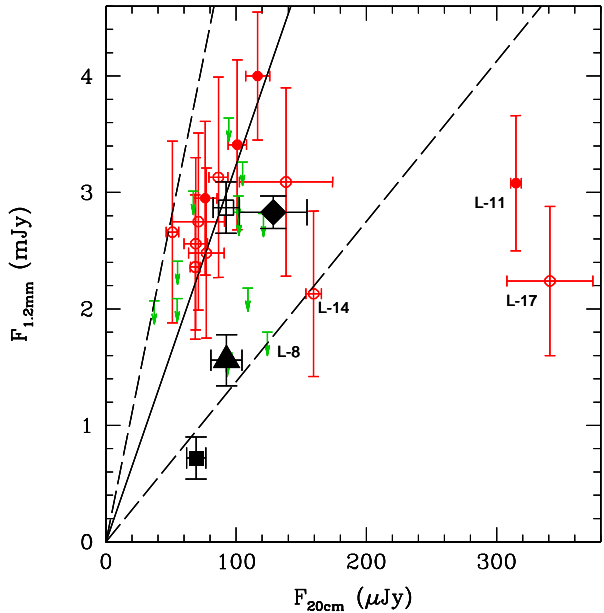}}
\caption{Observed MAMBO 1.2\,mm flux density as a function of 20\,cm flux density. The large black symbols show different stacked values: the entire sample (filled triangle),
  all sources with $> 3\sigma$ (filled diamond), the sources with 
$> 3\sigma$ signal without the two strongest radio sources (open square), and 
 the sources with $<$\,3\,$\sigma$ signal (filled square).  
 The filled (open) red circles are the sources that are  $4\sigma$ detections 
($3\sigma$ tentative detections) at 1.2\,mm. 
 The green arrows are 2\,$\sigma$ upper limits for the sources with 
S/N\,$<$\,3 at 1.2\,mm.  
The solid black line shows the correlation derived from \citet{Chap05} for $z\simeq 2$ (assuming $F_{\rm 850\,\mu m}/F_{\rm 1.2\,mm} = 2.5$ at $z\sim 2$).
 The dashed lines are the limits of this correlation at $\pm 1\sigma$.  The labelled sources are those with the highest 20\,cm/1.2\,mm flux density ratios (see Table \ref{res}).}
\label{1220.eps}
\end{figure}

The studied sources benefit from exceptionally deep VLA data at 1.4\,GHz
 \citep[rms\,=\,2.7\,$\mu$Jy in the center of the field, 12$-$15\,$\mu$Jy in
most of the 0.5\,deg$^{2}$ field;][]{OwMo09}. Such a depth yields
radio detections for almost the entire sample. 8 of 33 sources
 are not detected due to a loss of radio sensitivity in the
outer parts of the field, largely due to the decrease in primary beam sensitivity and bandwidth smearing.  
 The GMRT 610\,MHz observations are also very deep (rms\,=\,10\,$\mu$Jy),
cover almost the entire field, and detect all 33 sources. 
The VLA 324\,MHz observations reach a depth of rms\,=\,70\,$\mu$Jy and cover the
entire field \citep{Owen09}. They yield detections for 17 sources. The radio flux densities are listed in
Table~\ref{res}. The cross identification was made by comparing the radio
and the SWIRE positions.  Following the method of \citet{Ivis07} and \citet{Down86}, we have verified that all our sources have very reliable radio 
associations,
 with an average probability of spu\-rious association in 2$\arcsec$ of 
\textless\,$P$\,\textgreater\,=\,0.001. For \citet{Ivis07}, the association 
is reliable if $P \leq 0.05$.

Based on the correlation between $L_{\rm FIR}$ and the radio
luminosity in star-forming regions and in local starburst galaxies
\citep{Helo85,Cond92,Craw96,Sand96}, the radio and FIR luminosities are
expected to be linked as well at $z\sim2$ \citep{Ibar08}. We discuss this in Sec.\ 4.4, together with the derivation of $L_{\rm FIR}$. 
However, if this correlation is verified at $z\sim2$, one may expect a straightforward relation  between the radio and 1.2\,mm flux densities for starbursts with similar SEDs. 
To check that, we plot the 1.2\,mm flux density as a function of the
radio flux density at 20\,cm in Fig. \ref{1220.eps}.
 Because of the limited sensitivity at 1.2\,mm, we consider average values of 1.2\,mm flux density and radio flux density, for the entire sample,
 for the 13 sources with 1.2\,mm S/N\,$>$\,3, and for the 20 sources with 1.2\,mm S/N\,$<$\,3 (Table \ref{stack2}).
 In Fig. \ref{1220.eps}, we also
report the ratio of the average flux densities and $1\sigma$ dispersion 
found by~\citet{Chap05} \citep[see also e.g. ][]{Cond92,Smai02,Yun02,Ivis02} for
a sample of radio-detected sub-millimeter galaxies at $z\sim 2$. The mm/radio ratios for both the whole sample and the S/N\,$>$\,3 sources are reasonably compatible with the correlation between 
 mm/submm and 
radio fluxes found for $z\sim 2$ starbursts~\citep{Chap05}.  However, the ratio between average values of 1.2\,mm flux density and radio flux density, for the entire sample and especially 
 for the 20 sources  with 1.2mm S/N\,$<$\,3, might be slightly smaller compared to submm selected galaxies. This could be explained as an effect of the 24$\mu$m selection,  
  and by a greater AGN contribution and/or hotter dust (see Sec. 5.3 and 5.4 for a complete discussion).

In order to assess whether the radio emission observed in our sources
is associated with AGN or with star-forming regions, one may also consider the radio spectral shape, the rest-frame radio luminosity, and the radio morphology \citep[see e.g.][]{Bigg08,Seym08}. The
radio spectral index $\alpha$, where $F_{\nu}\propto \nu^{\alpha}$, is 
first calculated from the flux densities at 20\,cm, 50\,cm, and 90\,cm, when they are available, using the best power law fit between these three wavelengths,
 and is
reported in Table~\ref{res}. When only two radio fluxes are available, the index is just derived from the flux density ratio. Seven sources have no determined spectral
 radio index because of a lack of radio detections at 1.4\,GHz and 324\,MHz.

Most of our sources have a radio spectral index in the range $\sim-$0.4 to $-$1.2 (average $<\alpha>=-0.64\pm0.07$; median$\,=\,-0.74$). This is not very discriminating since, 
considering the uncertainties, such values of $\alpha$ are typical for star-forming galaxies, type II AGN, and many radio galaxies 
\citep[e.g.,][]{Cond92,Poll00,Cili03,Breu00,Breu01}.  
About 20\% of the sources, mostly among those with some radio excess, could either be in the same range or have $\alpha$\,$\gtrsim$\,-0.5 typical of flat spectrum sources.

The radio luminosity and the FIR-radio correlation  
are most often expressed in term of the rest-frame luminosity at 1.4\,GHz, 
$L_{\rm 1.4\,GHz}$. With the same assumptions as for the SMG sample of 
\citet{Kova06}, we can rewrite their Eq. (7) as

\begin{equation}
L_{\rm 1.4\,GHz}({\rm W\,Hz^{-1}})=4\pi D_{L}^{2}F_{\rm 20\,cm}
(1+z)^{(-\alpha-1)}
\label{lum_radio}
\end{equation}
\noindent where $D_L$ is the luminosity distance. 
In cases with detections at at least two radio wavelengths, we assume the derived value of $\alpha$. When only $F_{\rm 50\,cm}$ is available, we assume a fixed $\alpha$ value equal
 to the average value found for our sample, $<\alpha>=-0.64$

Using the flux at 610\,MHz, this equation becomes

\begin{equation}
L_{\rm 1.4\,GHz}({\rm W\,Hz^{-1}})=4\pi D_{L}^{2}F_{\rm 50\,cm}\left(\frac{2.3}{1+z}\right)^{\alpha}\frac{1}{1+z}.
\label{lum_radio50}
\end{equation}

Finally, we examine the radio sizes of our sources. A large size 
($\gtrsim1\arcsec$) may indicate radio jets or lobes
associated with the presence of a radio loud AGN. However, it is also possible that the extended radio emission is associated with star-forming regions distributed across the galaxy, as is the case for some local ULIRGs \citep[e.g.,][]{Murp01} and a large fraction of classical SMGs, e.g., $\sim$70\% in \citet{Chap04} \citep[see also ][]{Seym08,Rich07,Bigg08}. Nine sources are extended
(major axis $\gtrsim$\,5\,kpc) in the VLA 20\,cm images, with most sizes
$\gtrsim$\,10\,kpc, and the remainder are either unresolved or not detected (see
Table~\ref{res}). The radio luminosities of the nine extended sources range
from $\sim 10^{23.9}\,$W\,Hz$^{-1}$ to $\sim 10^{24.6}$\,W\,Hz$^{-1}$.

\subsection{Far-infrared luminosity}\label{lfir_sfr}

Estimates of $L_{\rm FIR}$ for our indivi\-dual 
sources are 
 quite uncertain due to the lack of data between 24\,$\mu$m and 1.2\,mm, where
most of the far-infrared e\-nergy is emitted. None of our
sources are detected at 70\,$\mu$m or 160\,$\mu$m at the SWIRE sensitivities
(the 3$\sigma$ limits at 70 and 160\,$\mu$m are 18 and 108\,mJy, respectively).
 There are not even detections in the $\sim 2.5$ times deeper MIPS images available from GO Program 30391 (PI: F. Owen) in the center of the field, which is well observed at 20\,cm. 
We thus make use of these deep MIPS data only for stacking the 70 and 160\,$\mu$m images to
constrain the average $L_{\rm FIR}$ of our sample.

We estimate average flux densities from the stacked MIPS images at 24, 70, and
160\,$\mu$m, following the method described in Lo09. An initial stack based
on the SWIRE MIPS images was made for the entire sample of 33 sources, but
it yields only a marginally significant ($\sim$2-3\,$\sigma$) detection at 160\,$\mu$m. The average flux densities at 70\,$\mu$m and 160\,$\mu$m are almost 5
times lower than the SWIRE 3\,$\sigma$ li\-mits (Table \ref{stack}). 
  We thus use stacks made with the  GO-30391 images of the 21 sources covered by these deeper MIPS observations. The results from stacks are
reported in Table~\ref{stack}. 
 In addition to the stacked flux densities for the
entire sample, we also report
stacked flux densities for two subsamples apiece: 10 sources
with 1.2\,mm SNR\,$>$\,3, and 11 sources with SNR\,$<$\,3.

The median stacked flux densities at 70\,$\mu$m and 160\,$\mu$m for the entire sample are
quite faint, about 3\,mJy and 13\,mJy respectively. We use these median stacked flux densities in the following analysis.
  There is no appreciable
difference in $<F_{\rm 160\,\mu m}>$ and $<F_{\rm 70\,\mu m>}$ between sources  tentatively detected ($>$3\,$\sigma$) and
undetected ($<$\,3$\sigma$) at 1.2\,mm. The median 160\,$\mu$m and 70\,$\mu$m flux densities seem smaller
than for the previous similar sample of \citet{Youn09}, which has higher average 24\,$\mu$m flux density (Table \ref{stack}).

We combine the median flux densities at 70\,$\mu$m, 160\,$\mu$m, and 1.2\,mm to
build a median far-IR SED for our sample. We fit the median FIR SED
assuming the average redshift of the sample ($z$\,=\,2.08) with a ``graybody'' 
model with a fixed value of the emissivity index \citep[ $\beta = 1.5$, see e.g. ][]{Kova06,Beel06} to derive 
the temperature, $T_{\rm dust}$, of a single dust component. The best-fit value for
the whole sample is $T_{\rm dust} = 37 \pm 8\,$K. This model yields $L_{\rm FIR} = 2.5 \times 10^{12}\,L_{\odot}$, which is well in the ULIRG
range and may be considered the average value for ``5.8\,$\mu$m-peakers''
with $F_{\rm 24\,\mu m} > 400\,\mu$Jy and $r > 23$. Note that these values for
$T_{\rm dust}$ and $<L_{\rm FIR}>$ are slightly smaller than the values of 
Lo09, $T_{\rm dust} = 41$\,K and $<L_{\rm FIR}> = 4.6 \times 
10^{12}\,L_{\odot}$ and of \citet{Youn09}, $T_{\rm dust} = 41\,$K and 
$<L_{\rm FIR}> = 3.8 \times 10^{12}\,L_{\odot}$.

The actual $L_{\rm FIR}$ of our sources will span some range around this  
value. Although it is not possible to derive accurate values of $L_{\rm FIR}$
for each source in the absence of individual detections at 160\,$\mu$m or at
another wavelength close to the FIR maximum of the SED, several alternative
approaches are possible for estimating individual $L_{\rm FIR}$.

One can simply infer $L_{\rm FIR}$ from the flux density at 1.2\,mm, 
$F_{\rm 1.2\,mm}$.
The value of $F_{\rm 1.2\,mm}$ can place strong constraints on $L_{\rm 
FIR}$ if we
assume that the whole $L_{\rm FIR}$ and the radiation detected at 1.2\,mm
are produced by dust heated by the same mechanism. The relation between
$L_{\rm FIR}$ and $F_{\rm 1.2\,mm}$ can be derived assuming a thermal spectral 
model.  A ``graybody'' model, with a single dust component and emissivity 
index, e.g. $\beta = 1.5$ \citep{Beel06}, yields:

\begin{equation}
L_{\rm FIR}(L_{\odot})=\kappa\times10^{12}\,F_{\rm 1.2\,mm}({\rm mJy})
\label{lfir_f12}
\end{equation}
\noindent where the proportionality factor, $\kappa \sim 1-5$, depends mainly on
$T_{\rm dust}$ but also by a factor of a few on redshift. For redshifts $\sim 2$, $\kappa = 1.3$ for $T_{\rm dust} = 35\,$K and $\kappa = 2.14$ for 
$T_{\rm dust} = 40\,$K.
The dust temperature,
$T_{\rm dust}$, is difficult to constrain without observations in the FIR$-$submm
range, e.g., at 160 or 350\,$\mu$m. $T_{\rm dust}$ is known to span a large range, from 21\,K to 60\,K in 350\,$\mu$m-detected
SMGs \citep{Kova06}, and from 34\,K to 47\,K in $z\sim 2$
\textit{Spitzer}-selected and mm-detected star-forming
gala\-xies \citep{Youn09}. In the absence of other information, we assume the
average values inferred from our stacks, $T_{\rm dust} = 37\pm 8$\,K
for the whole sample, and $36 \pm 3$\,K and $39 \pm 2$\,K for the 
samples with 1.2\,mm S/N\,$>$\,3 and $<$\,3, respectively.  
 Assuming a ``graybody''
model with $T_{\rm dust} = 37\,{\rm K}$, 36\,K and 39\,K, the values 
of the factor $\kappa$ are 1.57, 1.41, and 1.93, respectively.
 $L_{\rm FIR}$ values thus derived using
Eq.~\ref{lfir_f12}, $L_{\rm FIR,mm}$, are reported in Table~\ref{lum} for the 1.2\,mm $4\sigma$ detections and 3\,$\sigma$ tentative detections. For the other non-detections, 
 we report upper limits to $L_{\rm FIR}$.

We also estimate $L_{\rm FIR}$ of each source using the available radio
data and the well-known FIR-radio relation for local
starbursts~\citep{Cond92}. The original definition is based on the 60\,$\mu$m and 100\,$\mu$m flux densities, but following \citet{Saji08}, we have adopted the definition:

\begin{equation}
q={\rm log}\,\left(\frac{L_{40-120}}{L_{\odot}}\right)-{\rm log}\,\left(\frac{L_{\rm 1.4\,GHz}}{\rm W\,Hz^{-1}}\right)+14.03
\end{equation}
\noindent where $L_{40-120}$ is the integrated IR luminosity between 40\,$\mu$m and 120\,$\mu$m in the rest frame. We assume $L_{\rm FIR,mm} = L_{40-120}$. 

\begin{figure}[!htbp]
\resizebox{\hsize}{!}{\includegraphics{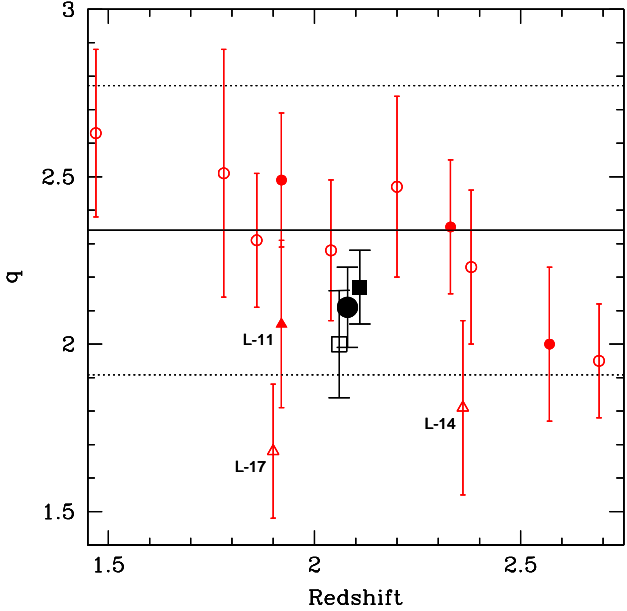}}
\caption{The radio-FIR $q$ factor (FIR-to-radio ratio) (Eq. 4). The large black symbols show the stacked values of different samples: the entire sample (33 sources, filled circle),
 the 13 sources with  S/N\,$>$\,3 at 1.2\,mm (filled square), and the 20 sources with S/N\,$<$\,3 (open square). 
 The solid black line shows the typical value $q = 2.34$ for local starbursts. The dotted black lines are the $3\sigma$ spread \citep{Yun01,Saji08}.
 $q$ values for individual sources are also shown for the sources with $> 4\sigma$ detections at 1.2\,mm (small filled symbols) and the sources with $> 3\sigma$ signal (small open symbols).
   The red triangles show the sources identified as having the highest 20\,cm/1.2\,mm flux density ratios.
 The red circles are the sources with no 20\,cm excess. 
}
\label{zq.eps}
\end{figure}

Although Eq.~(4) has been found to hold for local
sources, its validity seems to be confirmed also at the redshifts of our
sources~\citep{Kova06,Youn09}. The relation between $L_{\rm FIR}$ and the 20\,cm 
or
50\,cm flux density 
 is thus straightforward, if we assume the typical value $q = 2.34$ and 
$L_{\rm FIR,mm} = L_{40-120}$

\begin{equation}
L_{\rm FIR}(L_{\odot})=2.04\times10^{-12}L_{\rm 1.4\,GHz}({\rm W\,Hz^{-1}}).
\label{lfir_radio}
\end{equation}

However, this method is valid only if the radio emission is not significantly
affected by an AGN. This expression will give an upper limit to the
star-formation related $L_{\rm FIR}$ if an AGN contributes significantly to
the radio emission. The $L_{\rm FIR}$ derived using
Eq.~\ref{lfir_radio}, $L_{\rm FIR,radio}$, are reported in Table~\ref{lum}.

The values of these two estimates of $L_{\rm FIR}$ are within a factor of 3
for most of the sample. They range from $\sim 0.5$ to
$\sim 10 \times 10^{12}\,L_{\odot}$, confirming that almost all our
sources are ULIRGs with luminosities greater than $10^{12}\,L_{\odot}$. The
average $L_{\rm FIR}$ derived from the radio flux densities is $< L_{\rm 
FIR,radio}> = (4.14 \pm 0.53) \times 10^{12}\,L_{\odot}$. 
 Since this value may be  slightly overestimated because of AGN, it is consistent with $L_{\rm FIR}$ derived from fitting the median
FIR SED, $< L_{\rm FIR} > = 2.5 \times 10^{12}\,L_{\odot}$.

Alternatively, the radio-FIR relation (Eq. 5) can  
be applied to identify or test for the presence of AGN-driven radio activity in our sample (similarly to Sec. 4.3 but slightly more rigo\-rously). 
The agreement or deviation from this correlation can be easily
expressed by the value of the $q$-factor, as reported in Table \ref{lum}. 
We obtain an
average value $<\,q\,>\,=\,2.11\pm0.12$ for the whole sample.

This is
not very different from the average value $< q > = 2.34$ found for local
star-forming galaxies \citep{Yun01}. As shown in Fig \ref{zq.eps}, the $q$-factors of the 
 three sources identified as having the highest 20\,cm/1.2\,mm flux density ratios seem to be, on ave\-rage, lower than those of most other sources. 
 This is not surprising, since the $q$ factor is another way to quantify the radio excess.

\subsection{Stellar mass}

The IRAC-based selection of our sample corresponds to a rest-frame NIR
selection; therefore, if we ignore a possibly significant AGN contribution
at these wavelengths, we are directly sampling the stellar component (Lo09).
Applying the method developed by \citet{Bert04}, we have derived an
estimation of the stellar mass for our sources. As summarized in Lo09, this
derivation uses not only the IRAC data, but also optical and 24\,$\mu$m flux densities 
to perform a mixed stellar population spectro-photometric synthesis, and to 
constrain the extinction and the luminosity of young stellar popu\-lations
\citep{Bert04}. A redshift -- here photometric -- is also required. The stellar masses, initially computed with a $0.15\,-\,120$\,$M_{\odot}$ \citet{Salp55} initial mass function (IMF), are converted 
to a $0.10\,-\,100$\,$M_{\odot}$ IMF by applying a correction factor of 1.162 \citep{Bert03}. All the
stellar masses are shown in Table \ref{lum}. Their values range from 0.8 to
7.2$\times10^{11}$\,$M_{\odot}$, with a median value of
1.37$\times10^{11}$\,$M_{\odot}$, and an average value 1.77$\times$10$^{11}$\,$M_{\odot}$.

As discussed in Lo09, these values may be somewhat overestimated
for several reasons: our models may not fully take into account the TP-AGB
 contribution to infrared light \citep{Mara05}; they assume a Salpeter-like IMF, 
which yields higher
masses than other IMFs; and although the near-IR SED is dominated by stellar
emission, it is possible that an AGN component is also present. Therefore,
these estimates should  be considered as upper limits to the true stellar
masses, and may be overestimated by a small factor up to $\sim$2.

Using the same method, Lo09
found a median stellar mass
 $M_{\star}$\,=\,1.80$\times10^{11}$\,$M_{\odot}$ and an average value 2.16$\times$10$^{11}$\,$M_{\odot}$, about 30\% larger than for our
sample. This is consistent with the comparison of the average values of the
5.8\,$\mu$m flux density of both samples, which reflects the rest-frame luminosity at
1.6\,$\mu$m: 53\,$\mu$Jy for our sample and 77\,$\mu$Jy for that of Lo09.
Indeed, as discussed by Lo09, the direct comparison of the NIR rest-frame
luminosity at 1.6\,$\mu$m is probably the most consistent way to compare
stellar masses in different samples, in order to avoid the dependence on the various
methods applied to infer stellar masses from infrared fluxes. We have thus
estimated the rest-frame luminosities at 1.6\,$\mu$m, $\nu L_{\nu}$(1.6\,$\mu$m)  (Table \ref{lum}), by interpolating the
observed IRAC fluxes as in Lo09. As expected, the average rest-frame
luminosity at 1.6\,$\mu$m in Lo09, 4.1$\times10^{11}$\,$L_{\odot}$, is a factor of 1.6 larger than for our sample, 2.5$\times10^{11}$\,$L_{\odot}$ (Table \ref{lum}).
The sources in our sample are as luminous as, or slightly more luminous at 1.6\,$\mu$m by a factor $\sim$\,1.5 
 than, submm selected SMGs, and should also be $\sim$\,1.5 times more massive than classical SMGs assuming the same mass-to-light ratio of Lo09. 
Our sample also shows a mass-to-light ratio consistent with the radio galaxy sample detected at 24\,$\mu$m by \citet{Seym07}, who computed stellar
 masses directly from the luminosity at 1.6\,$\mu$m in the rest-frame.  

\subsection{Star formation rate}

\begin{figure}[!htbp]
\resizebox{\hsize}{!}{\includegraphics{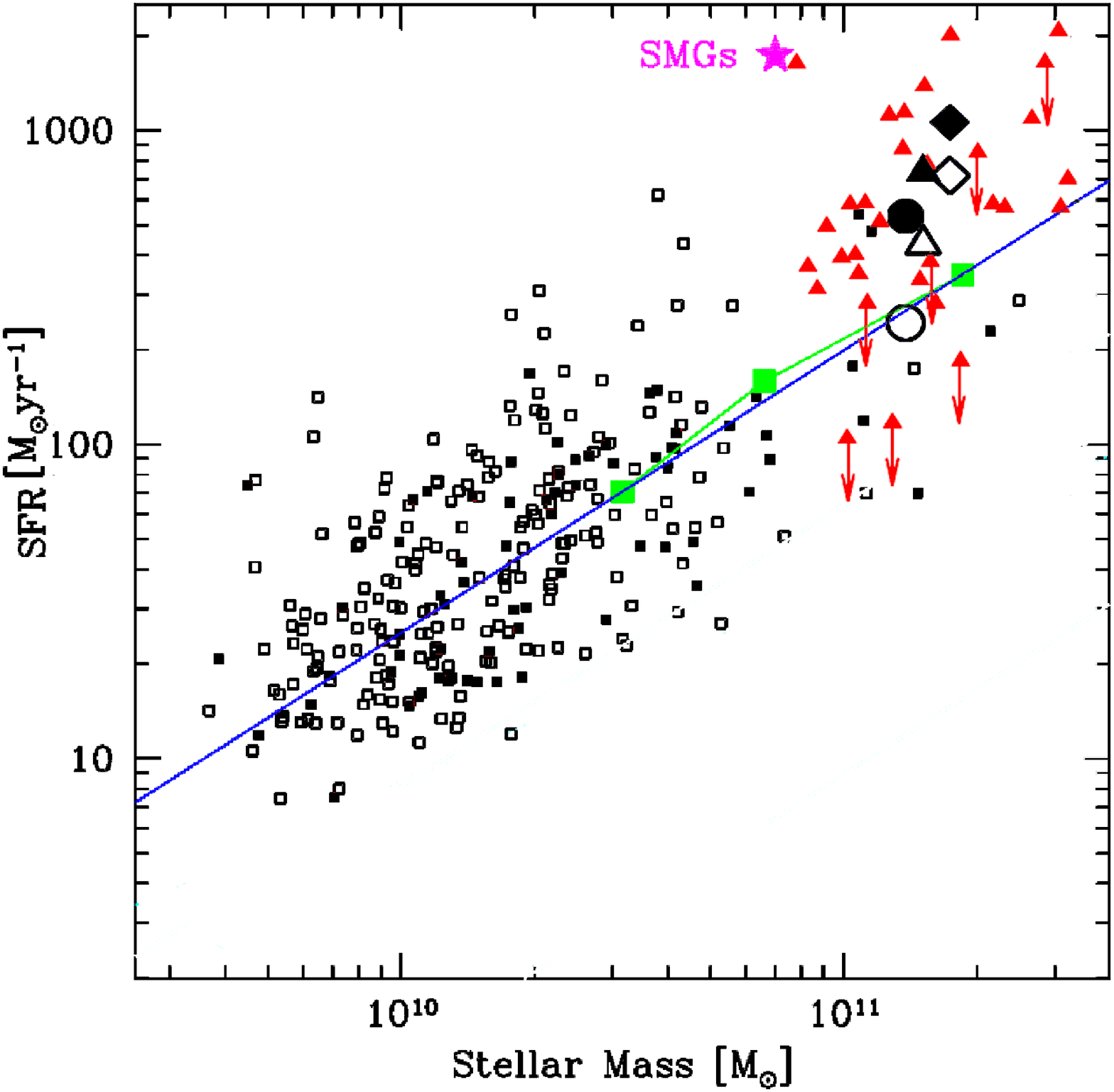}}
\caption{{\it \citep[Adapted from Fig.\ 14 of][]{Dadd07}}. Star formation rate vs. stellar mass. Values of SFR deduced from radio data for the 
 sources of our sample are represented by the red triangles (see Secs. 4.4, 4.6 and Table~\ref{lum}). The arrows are 2\,$\sigma$ limits. 
 The large black symbols show stacked values for different samples:  all sources (radio determination = filled triangle, millimeter determination = open 
triangle), 
 the sources with $>$3\,$\sigma$ signal at 1.2\,mm (radio determination = 
filled diamond, millimeter determination = open diamond), and the sources with 
$<$3\,$\sigma$ signal at 1.2\,mm (radio determination = filled circle, 
millimeter determination = open circle).
The small black squares are for the 24\,$\mu$m BzK sources from GOODS \citep{Dadd07}. The magenta star shows a typical value for SMGs. The large green squares trace the average SFR-mass relation
 in GOODS-N (160\,arcmin$^2$) determined from radio stacking of $K$\,$<$\,20.5 galaxies in three mass bins; the blue line is SFR\,=\,200\,$M_{11}^{0.9}$ ($M_{\odot}$\,yr$^{-1}$), where $M_{11}$ is 
the stellar mass in units of 10$^{11}$\,$M_{\odot}$ \citep{Dadd07}. }
\label{m_sfr.eps}
\end{figure}

From the estimated $L_{\rm FIR}$, we derive the star formation
rate (SFR) of our sources assuming the relation from~\citet{Kenn98} using a $0.10\,-\,100$\,$M_{\odot}$ Salpeter IMF:

\begin{equation}
\rm{SFR}({\textit M}_{\odot}\,yr^{-1})\approx1.8\times10^{-10}{\textit L}_{FIR}({\textit L}_{\odot}).
\end{equation}

The SFR values derived from $L_{\rm{FIR,mm}}$ and $L_{\rm{FIR,radio}}$ are listed in
Table~\ref{lum}. The SFRs derived from $L_{\rm{FIR,radio}}$ for the sources with possible radio excess may be overestimated due to the possible AGN contribution to
their radio emission.
The mean SFR derived by ave\-raging the radio-based estimates of all sources is
$\sim$750\,$M_{\odot}$\,yr$^{-1}$; the mean derived from the average $L_{\rm FIR}$ is $\sim$450\,$M_{\odot}$\,yr$^{-1}$.

To estimate the contribution of ``5.8\,$\mu$m-peakers'' with $F_{\rm{24\,\mu
m}}$\,$>$\,400\,$\mu$Jy to the star formation rate density (SFRD) of the
universe, we consider the space density and average SFR of our sources.
Assuming a redshift interval of 1.5$<z<$2.5, and an average SFR of
450\,$M_{\odot}$\,yr$^{-1}$, we derive the contribution of our sample to the
(comoving) SFRD to be $\sim$1.5$-$4$\times$10$^{-3}$\,$M_{\odot}$\,yr$^{-1}$\,Mpc$^{-3}$. This
value corresponds to $\sim$5\% of the SFRD of all classical SMGs
\citep{Aret07,Chap05}, i.e., to $\sim$10\% of the SFRD of SMGs in the
interval 1.5$<z<$2.5, and up to $\sim$15\% in the interval of 1.7$<z<$2.3
where ``5.8\,$\mu$m-peakers'' are mostly confined.

The specific star formation rate (SSFR), defined as SFR/M$_{\star}$, ranges from $\sim$10$^{-8}$\,yr$^{-1}$ to $\sim$10$^{-9}$\,yr$^{-1}$ for our sample. 
There is a tendency for the sources with the lowest stellar masses to have the highest SSFRs. 
This result is consistent with, e.g., \citet{Noes07}. 
However, it is seen in Fig. \ref{m_sfr.eps} that ``5.8\,$\mu$m-peakers'' have significantly higher SSFRs than classical sBzK galaxies of comparable masses on average (Sec. 5.3).

\section{Discussion}

\subsection{Comparison with other samples: 1) \textit{Spitzer} selection (see Table \ref{samples})}

\begin{figure}[!htbp]
\resizebox{\hsize}{!}{\includegraphics{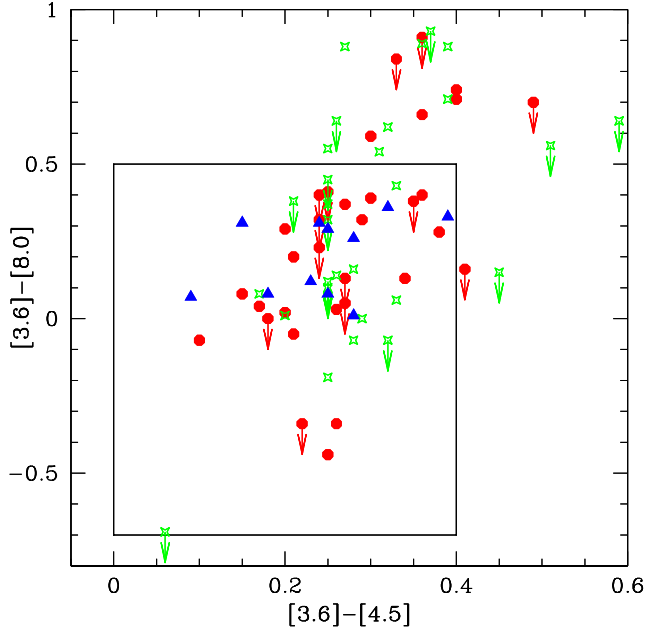}}
\caption{IRAC color-color diagram for several samples. The sample of \citet{Huan08} and \citet{Youn09} is represented by the solid blue triangles. The red circles show our sample. 
The open green stars are the sample of \citet{Lons08}.  The arrows are 3$\sigma$ upper limits for the difference [3.6]-[8.0]. 
The inset box shows the selection criteria of \citet{Huan08}. Magnitudes are on the AB scale.}
\label{cfa.eps}
\end{figure}

The present work is an extension of the study presented in Lo09 on a sample
of 61 sources selected in four SWIRE fields. The sources in Lo09 meet the same
selection criteria and span the same redshift range, 1.4\,\textless\,
$z_{\rm{phot}}$\,\textless\,2.7, as our sample. However, their selection is biased
toward brighter sources at 24\,$\mu$m, while the present sample is complete
to a 24\,$\mu$m flux density limit of 400\,$\mu$Jy. This bias results in an ave\-rage
24\,$\mu$m flux density a factor 1.5 greater than in our sample,
$<$\,$F_{\rm{24\mu m}}$\,$>$\,=\,819\,$\mu$Jy vs 566\,$\mu$Jy (see \S~\ref{sample}).
Nevertheless, the 1.2\,mm properties of the two samples are comparable. The
average flux densities at 1.2\,mm are similar: 1.56$\pm$0.22\,mJy for our
entire sample, against 1.49$\pm$0.18\,mJy for the sample in Lo09.

Although the source selection in Lo09 was optimized
to favor 1.2\,mm bright sources, their detection rate is slightly lower than
what we achieve with our complete sample, only 26\% (31\% in their best observed
field with a sensitivity similar to ours) with S/N$>$3\,$\sigma$ compared to 39\% in our sample. Both samples are characterized by a similar average
 $L_{\rm FIR}$ determined using a thermal spectral model (see Sec 4.4 and Eq.3), $<$\,$L_{\rm{FIR,mm}}$\,$>$\,=\,2.5$\pm$0.4$\times10^{12}$\,$L_{\odot}$ in this work
vs (2.8$\pm$0.1)$\times10^{12}$\,$L_{\odot}$ in Lo09 (restricted to the Lockman Hole
field). Thus, in spite of the refinement implemented by Lo09 in
their sample selection to increase the chance of finding mm bright sources
among ``5.8\,$\mu$m peakers'', the average mm pro\-perties of their sample and our
complete sample are consistent. 

\citet{Youn09} have studied the FIR properties of a similar \textit{Spitzer}-selected sample, based on MIPS 70 and 160\,$\mu$m detections and MAMBO 1.2\,mm observations. 
This sample \citep[see also ][]{Huan08} is selected based on IRAC colors, and
24\,$\mu$m flux densities (see Fig.~\ref{cfa.eps},  Table~\ref{samples} and \S~\ref{samples}), yielding 12
starburst galaxies with overall properties (redshifts, IRAC colors, 1.2\,mm
flux densities) similar to those in Lo09 and in this work. Consequently, the
 $L_{\rm FIR}$ and derived star formation rates are also similar in all these samples.

Finally, \citet{Lutz05} studied the mm properties of a \textit{Spitzer} sample
selected based on faint $R$ band magnitudes, relatively bright 24\,$\mu$m
flux densities ($F_{\rm{24\,\mu m}}$\,\textgreater\,1\,mJy), and high $F_{\rm{24\,\mu m}}$/$F_{\rm{8\,\mu
m}}$ ratios~\citep{Yan05,Yan07,Saji08}. This sample contains 40
sources at $z\sim$2 exhibiting both starburst and AGN properties. They are
on average fainter 1.2\,mm emitters than our sources and they show
significantly lower $F_{\rm{1.2\,mm}}$/$F_{\rm{24\,\mu m}}$ flux density ratios, with 
$<$\,$F_{\rm{1.2\,mm}}$\,$>$\,=\,0.63$\pm$0.20\,mJy (simple average) and $<$\,$F_{\rm{1.2\,mm}}$\,$>$/$<$\,$F_{\rm{24\,\mu m}}$\,$>$\,=\,0.43 (simple averages), compared to 
$<$\,$F_{\rm{1.2\,mm}}$\,$>$\,=\,1.56$\pm$0.22\,mJy and $<$\,$F_{\rm{1.2\,mm}}$\,$>$/$<$\,$F_{\rm{24\,\mu m}}$\,$>$\,=\,2.76$\pm0.50$ for our sample (Table \ref{stack2}; see also Fig.~\ref{1224z1.eps}). These differences are
due to the fact that in the majority of these objects, the main source of
power is an AGN rather than the powerful starburst required to produce bright mm
flux, with the AGN dominating their 24\,$\mu$m
emission.

\subsection{Comparison with other samples: 2) Submillimetre selection}

\begin{figure}[!htbp]
\resizebox{\hsize}{!}{\includegraphics{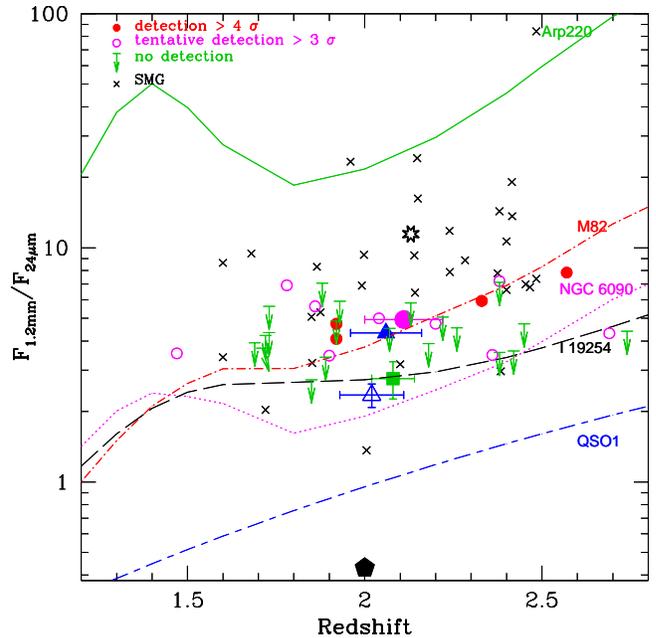}}
\caption{Observed flux density ratio, $F_{\rm{1.2\,mm}}$/$F_{\rm{24\,\mu m}}$, versus redshift. The small filled (open) magenta circles are the $4\,\sigma$ detections (tentative $3\,\sigma$ detections) at 1.2\,mm. The green arrows show the $2\,\sigma$ upper-limits for the other non-detections. 
The crosses show values for a sample of SMGs (see text). 1.2\,mm flux densities for this sample were derived from $F_{\rm{850\,\mu m}}$ when $F_{\rm{1.2\,mm}}$ is not available. 
 The large symbols are the  ratio of average flux density vs average redshift for different samples: all 33 of our sources (filled green square); all 13 of our sources 
with a $> 3\sigma$ signal (magenta circle);   the entire sample in Lo09 restricted to the Lockman Hole field (open blue triangle); 
 the sources with signal $> 3\sigma$ in Lo09 (filled blue triangle); 
 the AGN-dominated 24\,$\mu$m bright sample of \citet{Lutz05} (black pentagon); and the SMG literature sample (open star) \citep[see Lo09;][]{Copp06,Ivis07}. 
Expected values for various starburst and AGN templates are also shown (Lo09): starbursts Arp\,220 (solid green line), M\,82 (dot-dashed red line), and 
NGC\,6090 (dotted magenta line); 
AGN-starburst composite IRAS19254$-$7245 South (dashed black line); and AGN 
``QSO1'' (short-long-dashed blue line).}
\label{1224z1.eps}
\end{figure}

In order to characterize more quantitatively the difference in $F_{\rm{1.2\,mm}}$/$F_{\rm{24\,\mu m}}$ flux density ratios between our sample of ``5.8\,$\mu$m-peakers'' and classical
SMGs, we show them as a function of redshift in
Fig. \ref{1224z1.eps}. The sample of classical SMGs is the sub-set of the
sample used in Lo09 completed by the SHADES SMG sample from~\citet{Copp06} and \citet{Ivis07} detected at 24\,$\mu$m with 1.5$<$$z$$<$2.5. We also display a sub-set of ``5.8\,$\mu$m-peakers'' from Lo09.
 Note that most of the redshifts available for the Lo09 and the
SHADES samples are photometric~\citep{Aret07}. We also show in
Fig. \ref{1224z1.eps} the expected flux ratios for representative starburst
and AGN templates. To facilitate the comparison, we also show the ratio of the average flux densities for the samples 
of ``5.8\,$\mu$m-peakers'' from this work and from Lo09, and for a sample of classical SMGs \citep[the SMG sample of Lo09 augmented by those of SHADES:][]{Copp06,Ivis07}. 
Our complete sample
 of ``5.8\,$\mu$m-peakers'' confirms that the ratio of the average
1.2\,mm and 24\,$\mu$m flux densities for this class of sources is 
smaller than for SMGs. However, the difference is slightly smaller than in
Lo09 because of our lower mean 24\,$\mu$m flux density. The $F_{\rm{1.2\,mm}}$/$F_{\rm{24\,\mu m}}$ flux density ratios of our sample are also similar to one for 
the lensed LBG cB58, despite the fact that this lensed 
LBG is an order of magnitude less luminous than the average of our sample \citep{Sian08}.

We have shown that ``5.8\,$\mu$m-peakers'' are $z\sim$2 ULIRGs and that about 40\%
of them are bright mm sources. Thus,  $\sim40\%$ of them also belong to the class
of SMGs (Table \ref{lum}), one of the main classes of high-$z$ ULIRGs. A detailed comparison
between the MIR and FIR properties of a large sample of
SMGs~\citep{Grev04,Pope05,Chap05,Bory05,Fray04,Hain09} and ``5.8\,$\mu$m-peakers''
is presented in Lo09. In this study, Lo09 find that most 
``5.8\,$\mu$m-peakers''
represent a sub-class of SMGs. The differences found by Lo09 with respect to
classical SMGs are mainly related to the selection criteria for the
``5.8\,$\mu$m-peakers''. More specifically, the \textit{Spitzer} selection
favours sources with redshifts mostly concentrated in the range
$z\,\sim$\,1.7-2.3, rather higher stellar masses than classical SMGs at
similar redshifts, brighter 24\,$\mu$m flux densities, and thus higher
$F_{\rm{1.2\,mm}}$/$F_{\rm{24\,\mu m}}$ flux density ratios (see Fig.~9 in Lo09 and
Fig.~\ref{1224z1.eps}), and likely warmer dust temperatures. The
mm-detected ``5.8\,$\mu$m-peakers'' and classical SMGs show similar mm/submm flux
densities, implying that the main difference in the 1.2\,mm/24\,$\mu$m ratio
comes from the 24\,$\mu$m rather than the 1.2\,mm intensity. This
result implies that a high 24\,$\mu$m flux density does not directly translate into
a high 1.2\,mm flux density, and thus into high $L_{\rm{FIR}}$ and SFR, in starburst galaxies
at $z\sim$2~\citep[see also][]{Pope08b,Riek08}.

\subsection{Nature of the sources: 1. Powerful starburst activity}

\begin{figure}[!htbp]
\resizebox{\hsize}{!}{\includegraphics{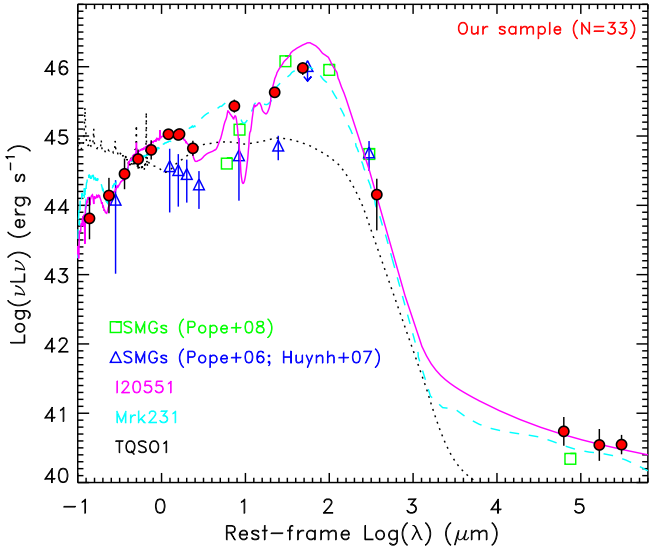}}
\caption{Median optical through radio SED of all 33 ``5.8\,$\mu$m-peakers''
 in our sample (red full circles).
The 70\,$\mu$m and 160\,$\mu$m median values
corres\-pond to the stacked values of the central 21 sources (see
\S~\ref{lfir_sfr}).  The magenta and cyan curves represent templates of
IRAS\,20551$-$4850 and Mrk\,231 normalized to the rest-frame luminosity at
1.6\,$\mu$m. The black dotted curve represents a type 1 AGN template
normalized at the lowest NIR data point (rest $\lambda$\,=\,2.4\,$\mu$m) of the median SED \citep{Poll07}. The blue
triangles represent the average fluxes of the sample of 31 HDFN
SMGs~\citep{Pope06}. The blue triangles at 70\,$\mu$m and 160\,$\mu$m
correspond to the stacked fluxes for a subsample of 26 sources from the same
sample~\citep{Huyn07}.
 The green open squares delineate the average SED of the
GOODS SMGs derived by~\citet{Pope08a}. }
\label{avg_sed} 
\end{figure}

As discussed in Sec. 2, our selection criteria are devised to distinguish
starburst ULIRGs from AGN through the presence of the redshifted
1.6\,$\mu$m stellar ``bump'' which dominates over a strong AGN continuum in
the near infrared. As shown by similar studies such as \citet{Weed06b}, 
\citet{Farr08}, \citet{Huan08}, and Lo09, such criteria are very successful in
selecting a majority of starbursts with strong mid-IR PAH features, and we
have good evidence that this is also the case for the present sample from the
combination of millimetre and radio data. 

Although there might be some fraction of ``5.8\,$\mu$m-peakers'' that
host a certain level of AGN activity, in the majority of these sources, the
optical to far-IR light is likely dominated by starburst emission. This is
confirmed by 1) the high detection rate at 1.2\,mm and the high average
value \,$<F_{\rm{1.2\,mm}}>$\,=\,1.56\,mJy;
and 2) the correlation between the mm and radio emission for the
majority of the sources.  
With star formation rates ranging from a few 10$^{2}$\,$M_\odot$\,yr$^{-1}$
to $\sim$10$^{3}$\,$M_\odot$\,yr$^{-1}$ and a mean
$<$SFR$>$\,$\gtrsim$\,450$-$750\,$M_\odot$\,yr$^{-1}$, they are powerful ULIRGs
with starburst strengths similar to those of SMGs.

Significant emission in the mid-IR bands of PAHs is a common feature in
starbursts \citep[see, e.g.,][ for local galaxies]{Rigo99,Desa07,Veil09,Farr09}. Although we are still lacking mid-IR spectroscopy for our $z\sim$2
sample to infer the precise strength of the objects' PAH emission, there are very good
reasons for thinking that it is strong and comparable to that observed in other
24\,$\mu$m-bright starburst galaxies at $z\sim$1.5$-$2 satisfying similar
criteria for selection of ``4.5\,$\mu$m-'' or ``5.8\,$\mu$m-peakers'' \citep[][ Lonsdale et al.\ in prep. ]{Weed06b,Yan07,Murp08,Huan08} or in about 30 ``4.5\,$\mu$m-peakers''
\citep{Farr08}. The strong PAH emission displayed in these objects
accounts for the major part of the mid-IR emission in the range
$\sim$6$-$12\,$\mu$m. It is thus clear that the PAHs generally contribute much of
the flux detected in the \textit{Spitzer}/MIPS broad 24\,$\mu$m band in such
``5.8\,$\mu$m-'' or ``4.5\,$\mu$m-peakers''. As noted, most SMGs have a much weaker 24\,$\mu$m intensity than our sources. 
It is not yet well understood why ``5.8\,$\mu$m-'' and ``4.5\,$\mu$m-peakers'' happen to have stronger PAH
emission without a pa\-rallel enhancement of the observed mm/submm flux
density, resulting in a higher 24\,$\mu$m/1.2\,mm ratio than the bulk of the
SMGs (Lo09).  \citet{Farr08} suggest the
possibility that star formation is extended on spatial scales of 1$-$4\,kpc
in such galaxies. 

In Figure~\ref{avg_sed}, we show the median SED of our
sample of ``5.8\,$\mu$m-peakers'', compared to  the templates of a starburst, an obscured
AGN, and an unobscured AGN. 
 It is clear that the median observed SED is much better
 matched by the starburst template than by the obscured AGN SED, where the 5.8\,$\mu$m bump is absent.

\begin{figure}
\resizebox{\hsize}{!}{\includegraphics{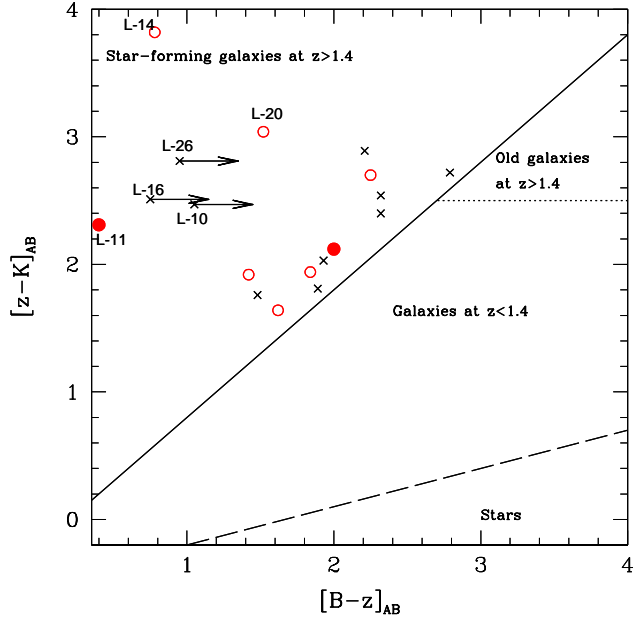}}
\caption{BzK color-color plot for our sample. $B$ magnitude is extrapolated from $u$ and $g^{\prime}$ magnitudes. When $z$ magnitude is missing, it is extra\-poled from $i^{\prime}$ and $J$. 
 Same symbols as Fig. \ref{1224.eps}. The black crosses are the sources with S/N\,$<$\,3 at 1.2\,mm.  The solid black line shows the selection criterion for sBzKs of \citet{Dadd04}. 
}
\label{bzk.eps}
\end{figure}

\begin{figure}
\resizebox{\hsize}{!}{\includegraphics{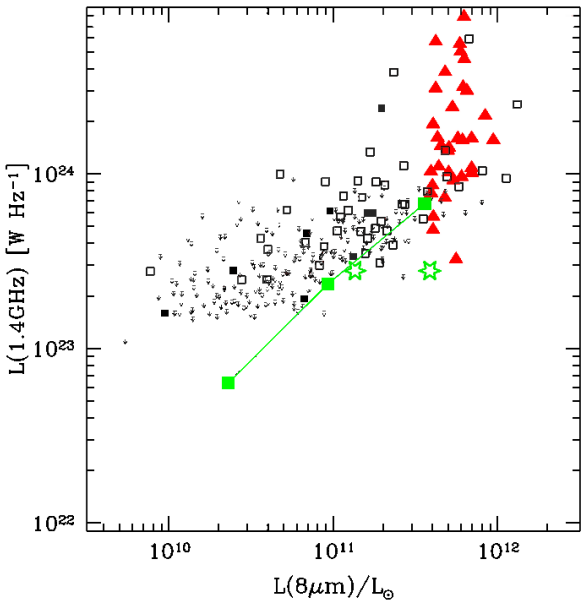}}
\caption{{\it \citep[Adapted from Fig.\ 7 of][]{Dadd07}}. Luminosity at 1.4\,GHz rest frame vs. luminosity at 8\,$\mu$m rest frame. ``5.8\,$\mu$m-peakers'' 
from our sample are displayed as large red triangles. 
Data for all 24\,$\mu$m BzK sources of the GOODS-N field  \citep{Dadd07} are shown for comparison as small black  symbols; the three large green squares show the average trend vs. 8\,$\mu$m luminosity, including both radio-detected and radio-undetected sources from \citet{Dadd07}.
The green stars are the two BzK sources with CO detections from \citet{Dadd08}.}
\label{L8.eps}
\end{figure}

``5.8\,$\mu$m-peakers'' (and ``4.5\,$\mu$m-peakers'') also have some relationship with other broader classes of infrared selected high-$z$ 
star-forming galaxies,
 as judged by 24\,$\mu$m intensity. As already discussed (e.g., Table 3), there are various samples \citep[e.g.][]{Magl07,Yan05,Houc05,Murp08}
 just selected as relatively strong at 24\,$\mu$m and weak in the optical, especially for the purpose of IRS mid-IR spectroscopy.
 One recent example is the so called ``Dust-Obscured Galaxies'' (DOGs) defined as having $S_{\rm{24\,\mu m}}$/$S_{\rm{R}}$\,$>$\,1000,
 with various limits for $S_{\rm{24\,\mu m}}$ such as 300\,$\mu$Jy by \citet{Dey08,Dey09}, or 100\,$\mu$Jy by \citet{Pope08b}. 
However, such broad criteria select a mixture of starbursts and AGN, especially those with power-law IRAC SEDs. 
 As seeing from Table 1, practically all sources of our sample satisfy or are close to the  $S_{\rm{24\,\mu m}}$/$S_{\rm{R}}$\,$>$\,1000 flux density 
ratio defining DOGs. 

A better established class of high-$z$ galaxies partially selected from optical-NIR colors is the ``BzK'' galaxies \citep{Dadd04}. 
This includes an ``sBzK'' sub-class of star-forming galaxies and a ``pBzK'' sub-class of passively evolving proto-ellipticals. 
 Fig.~\ref{bzk.eps} shows that all sources in our sample with $B$, $z$, and 
$K$ information are sBzK galaxies. 
``5.8\,$\mu$m-peakers''  share the same redshift range as the bulk of BzKs \citep[see e.g. Fig.\ 2 of][]{Dadd07}. 
However, they are at the very top of the BzK luminosity function. This is clearly shown in Fig.~\ref{L8.eps} where we have put our sample of 
``5.8\,$\mu$m-peakers'' 
on top of Fig.\ 7 of \citet{Dadd07}, which displays the rest luminosity $L_\nu$(8$\mu$m) (roughly proportional to the bulk of $S_{\rm{24\,\mu m}}$) and $L_{\rm{IR}}$ 
(deduced from $L_{\rm{1.4\,GHz}}$) for all BzKs in the GOODS-N field. 

We note that the two $z\sim$1.5 BzKs where CO was recently detected by \citet{Dadd08} 
are ``4.5\,$\mu$m-peakers'' with $S_{\rm{24\,\mu m}}$\,=\,140 and 400\,$\mu$Jy and $L_{\rm{IR}}$ close to 10$^{12}$\,$L_\odot$, i.e., comparable to or below 
the lower end of the luminosity distribution of our sample (Fig. \ref{L8.eps}). This probably means that CO should be easily detectable in most of our ``5.8\,$\mu$m-peakers''
with the current sensitivity of the IRAM Plateau de Bure Interferometer (PdBI), as shown by Yan et al. (in prep.), who detected strong CO emission in two ``5.8\,$\mu$m-peakers'' of \citet{Saji08}.

Since sBzK sources are major contributors to star formation at $z\sim$2, it is interesting to compare their star formation properties with those of our ``5.8\,$\mu$m-peakers''. 
 Figure~\ref{m_sfr.eps} shows that the average SFR of ``5.8\,$\mu$m-peakers'' is significantly greater than the average for sBzKs at $z\sim$2 \citep{Dadd07}. 
Even their SSFRs are significantly higher than those of sBzKs with comparable masses. This figure confirms that our ``5.8$\mu$m-peakers'' represent as significant a fraction
 of the most massive and star-forming sBzKs as Fig. \ref{L8.eps} implies.
On the other hand, as the average SFR of ``5.8\,$\mu$m-peakers'' is significantly lower than those of the bulk of classical SMGs, and their stellar mass slightly higher,
 it is not surprising that the mean SSFR of ``5.8\,$\mu$m-peakers'' is markedly smaller than one of SMGs (Fig.~\ref{m_sfr.eps}).

Another class of powerful starburst galaxies are the so called ``submillimetre faint radio galaxies'' \citep[SFRGs;][]{Blai04,Chap04,Chap08}. 
They are defined as radio sources with radio fluxes similar to those of SMGs which are not detected in typical SCUBA surveys. In addition to AGN, this class of sources
may contain star-forming galaxies characterized by slightly hotter temperatures
than typically observed in submillimetre galaxies. It is also possible that
the radio pro\-perties and the hotter dust temperatures in these sources might
be due to the presence of AGN activity, but clear evidence is still
lacking for the relative fractions of AGN and hotter starbursts~\citep{Chap08,Case08,Case09}. 
It is thus interesting to consider whether our sources undetected at 1.2\,mm and lying outside or at the outskirts of the radio$-$1.2\,mm correlation might belong to the class of SFRG. 
More precisely, \citet{Chap08} propose to reserve the name of SFRG for radio sources with 
$L_{\rm{1.4\,GHz}}$ \,$>$\,$10^{24}$\,W\,Hz$^{-1}$. 
As SFRGs must also not be detected in 850\,$\mu$m surveys with typical sensitivity  4\,mJy, we assume that SFRGs in our sample may be defined 
by $L_{\rm{1.4\,GHz}}$ \,$>$\,10$^{24}$\,W\,Hz$^{-1}$ and $F_{\rm{1.2\,mm}}$\,$<$\,1.6\,mJy. SFRGs are identified with these criteria in Table 5. 
There are 3 confirmed SFRGs and 3 tentative SFRGs (together with 4 confirmed SMGs and 9 tentative SMGs defined as satisfying the limit $F_{\rm{1.2\,mm}}$\,$>$\,1.6\,mJy,
 and 7 sources too weak in radio to be SFRG and at 1.2\,mm to be SMG). We may thus estimate that among ``5.8\,$\mu$m-peakers'',
 at least $\sim$40\% are SMGs, $\sim$10-20\% may be SFRGs, and $\sim$15-20\% are weak radio and mm sources. 

\subsection{Nature of the sources: 2. Weak AGN activity}

Our selection criteria were designed to select
starburst ULIRGs over AGN through the presence of the redshifted
1.6\,$\mu$m stellar ``bump''. However, similar studies have shown that the discri\-mination against AGN with such
criteria is not perfect and that a small fraction  
has a weak but significant AGN contribution in
their mid-IR SED. Since AGNs are power\-ful NIR emitters due to their ability 
to heat dust up to its sublimation temperature, we expect that such emission may
smear out the 
1.6\,$\mu$m stellar ``bump'', flatten it, or move to longer observed wavelengths \citep{Bert07b,Dadd07}. However, some AGNs can satisfy our selection criteria. Indeed, about 8 radio-galaxies 
from the sample of \citet{Seym07} follow our criteria. But these sources show a significant radio excess \citep{Arch01,Reul04} due to the AGN.

\begin{figure}[!htbp]
\resizebox{\hsize}{!}{\includegraphics{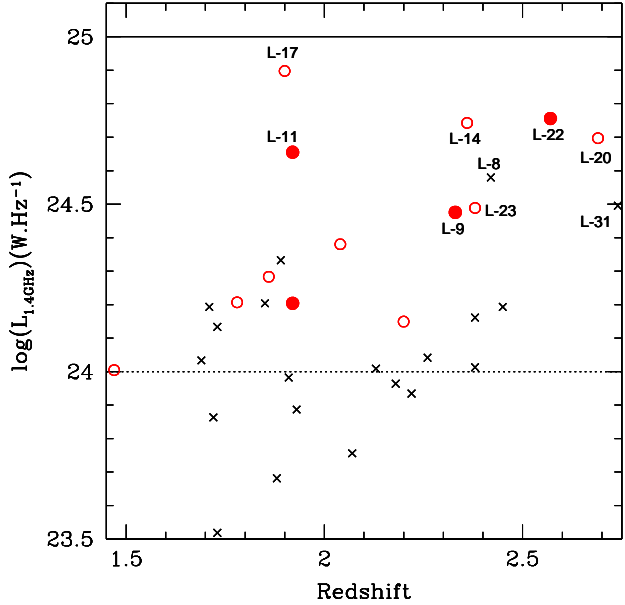}}
\caption{1.4\,GHz luminosity as function of redshift. Same symbols as Figs. \ref{1224.eps} and \ref{bzk.eps}.   
 The solid black line shows the radio loudness limit for 1.4\,GHz luminosity \citep{Saji07}.}
\label{zlum.eps}
\end{figure}

To identify, quantify, and characterize any AGN activity which might be
present in our sample it would be useful to have mid-IR spectroscopic
data~\citep[see, e.g.,][]{Farr08,Weed06b}, optical/NIR
spectroscopic data~\citep{Chap05,Bert07a}, morphological data from high
angular resolution radio observations~\citep[see, e.g.,][]{Rich07,Muxl05,Bigg08}, or
deep X-ray data~\citep[see, e.g.,][]{Alex05}. Only limited observations
of this kind are available for our sample. A spectroscopic observation from
Keck/LRIS is available for source L-25~\citep[LH\_574364
in][]{Bert07a}. Based on one emission line identified as MgII, this source
is classified as a type 2 AGN, although its radio/1.2mm flux density ratio is not one of the highest and its radio spectral index instead point to a starburst-dominated IR SED (Table \ref{res}).

X-ray data of moderate depth (70\,ks exposures)
from \textit{Chandra} are also available for all our sources~\citep{Poll06}.
None of them is detected by \textit{Chandra} to a 0.3$-$8\,keV flux
limit of 3$\times$10$^{-15}$\,erg\,cm$^{-2}$\,s$^{-1}$. We stacked the X-ray
images of all 33 sources, but no significant detection was obtained. We can
only set an upper limit to the average 0.3$-$8\,keV flux of
10$^{-16}$\,erg\,cm$^{-2}$\,s$^{-1}$ which corresponds to an average X-ray
luminosity of 2$\times$10$^{42}$\,erg\,s$^{-1}$. Such a low signal does not
allow us to make any claim about the AGN contribution and X-ray properties of 
our sample. 

The multi-frequency radio observations available for the majority of our
sources may also allow us  to investigate the AGN contribution. As discussed in Sec. 4.3, we can find
some indication of AGN activity based on the 1.2\,mm/radio
flux density ratio complemented by  the spectral index and the radio size; 
however, an extension may as easily reflect well an extended starburst as an AGN lobe.

Two 20\,cm radio sources, L-17 and L-11, are specially strong  ($F_{\rm{20\,cm}}\,>$\,300\,$\mu$Jy).
The 20\,cm flux density of L-17 is slightly greater than 300\,$\mu$Jy,
implying a 1.4\,GHz luminosity close to 10$^{25}$\,W\,Hz$^{-1}$ (see Fig.
\ref{zlum.eps}) which is close to the 
radio loudness limit at $z\sim$2~\citep{Jian07, Saji07}. L-11, despite
its high 20\,cm flux, is not strictly radio loud according to this
defi\-nition. Neither source is resolved in the radio; however, this
might be not very meaningful, since both sources are in regions with limited
sensitivity at 20\,cm. Their spectral indices are $-$0.76 and
$-$0.45,  respectively, which is not very discriminating.  
 Interestingly, both sources are rather strong and well
detected at 1.2\,mm (S/N\,$=$\,3.5 and 5.3 respectively), implying that, if they are  AGN,  they might experience both
starburst and accretion activity.

The two other sources with  evidence of possible AGN-driven radio activity based on
a high 20\,cm/1.2\,mm flux density ratio, L-8 and L-14,  
 are not detected at 1.2\,mm (S/N\,$<$\,1 and S/N\,$=$\,3, respectively), L-8 could be an SFRG, and L-14 either an SMG or an SFRG. L-8 is also extended in its radio image.

\section{Summary and conclusion}

The aim of this project was to determine the average properties of a 
 complete 24\,$\mu$m flux limited sample of bright \textit{Spitzer} sources selected
 to be starburst dominated at $z\sim$2, using multi-wavelength data. 
The sample of 33 $z\sim$2
SMG candidates was built with all the optically faint sources in a $\sim$0.5\,deg$^{2}$ area
of the Lockman Hole SWIRE field meeting selection criteria based on
MIPS/IRAC fluxes. These criteria are 
$F_{\rm{24\,\mu m}}$\,\textgreater\,400\,mJy, a peak in the 5.8\,$\mu$m IRAC band due
to redshifted 1.6\,$\mu$m stellar emission, and $r^{\prime}_{\rm{Vega}}$\,$>$\,23. The
J1046+59 field was selected because of the availability of very deep radio observations at
20\,cm and 90\,cm with the VLA and at 50\,cm with the GMRT.
 All sources in our sample are detected at 50\,cm.

The entire
sample has an average 1.2\,mm flux density of 1.56$\pm$0.22\,mJy. However, the limited sensitivity allowed only four confirmed 4\,$\sigma$ detections, plus nine tentative 3\,$\sigma$ detections.
Since the average 1.2\,mm flux density, 1.56\,mJy, corresponds to a 850\,$\mu$m flux density close to 4\,mJy, about half of the sources
may be considered SMGs. However, their redshifts range from $z\sim$1.7$-$2.3, similarly to the sample in \citet{Huan08}, but smaller than the redshift range covered by SMGs. 
The sample selected here is characterized by brighter
24\,$\mu$m flux densities, on average, than those of SMGs, and consequently shows 
systematically lower $F_{\rm{1.2\,mm}}$/$F_{\rm{24\,\mu m}}$ 
ratios than classical
SMGs. It is quite likely that our selection favours the SMGs with the
brightest 24\,$\mu$m flux densities, due probably to enhanced PAH emission.

From stacking individual images of the sources, we are able to build the
median FIR SED of our sample and estimate the corresponding $L_{\rm{FIR}}$, SFR and
$T_{\rm{dust}}$ assuming a single temperature ``greybody'' model. The inferred values
are $T_{\rm{dust}}$\,=\,37$\pm$8\,K, $L_{\rm{FIR}}$\,=\,2.5$\times$10$^{12}$\,$L_{\odot}$, and SFR\,=\,450\,$M_{\odot}$\,yr$^{-1}$. These estimates indicate that most of the
sources are ULIRGs. However, estimates of $L_{\rm{FIR}}$ for individual sources deduced
from the IR-mm SED are highly uncertain due to the lack of flux measurements
between 100\,$\mu$m and 500\,$\mu$m. The high quality radio data provide
important complementary information on $L_{\rm FIR}$ and the star
formation rate, since the 1.2\,mm/radio flux density ratio of the majority of
individual sources is consistent with the FIR/radio correlation, which allows a
derivation of $L_{\rm{FIR}}$ and SFR from the radio flux, providing further
confirmation that most of the selected sources are ULIRGs. The average value of  $L_{\rm{FIR}}$  inferred from the FIR-radio correlation is 4.1$\times$10$^{12}$\,$L_{\odot}$;  
  however, this value may be overestimated because of an AGN contribution.

Stellar masses are estimated by modelling the optical-IR SED with stellar
population synthesis models. They are of order a few 10$^{11}$\,$M_{\odot}$.
Roughly scaling with the observed 5.8\,$\mu$m fluxes, they are similar to those of 
other samples of 24\,$\mu$m-bright, $z\sim$2 \textit{Spitzer} starbursts, and
slightly higher than those of classical SMGs.

Overall, this sample appears similar to other samples of
\textit{Spitzer} $z\sim$2 SMGs~\citep[Lo09;][]{Youn09} in terms of 
millimetre emission, $L_{\rm FIR}$, and SFR.

The complete radio detection of
all sources provides a good estimate of the total star formation rate
of such sources. They represent a significant fraction of all SMGs in the
redshift range $z\sim$\,1.7-2.3 ($\sim$10$-$15\%). 
 Most of these ``5.8\,$\mu$m-peakers'' are star-forming BzK galaxies with luminosities at the top of the luminosity distribution of sBzKs. 

The surface density of ``5.8\,$\mu$m-peakers'' has been found to be  61\,deg$^{-2}$ by \citet{Farr06}.
This is consistent with 33 sources in the 0.49\,deg$^{2}$ of our field. We may thus estimate that 40\,$-$\,60 similar ``5.8$\mu$m-peakers'' per square degree  
could be identified in the full SWIRE survey (49\,deg$^{2}$).
Most of them should be $z\sim2$ starburst ULIRGs. At least half of them may be considered to be SMGs, including a small fraction of composite obscured AGN/starburst objects.
 Another significant fraction may be considered as SFRGs.

These results illustrate the power of deep multi-$\lambda$ studies, especially with complete radio data, for analysing populations of powerful high-$z$ IR and submm sources.
 Such deep radio data are essential for disentangling starbursts and infrared-bright AGN, and for easily providing estimates of their star formation rates. 
We note especially the impressive complete detection of relatively weak $z\sim$2 SMGs over 0.5\,deg$^2$ in a single pointing of the GMRT at 610\,MHz. 
As already proved by the analysis of SCUBA sources \citep[e.g.,][]{Ivis02}, radio data are essential for identifying optical/near-IR counterparts and analysing submm surveys. 
This will be even more crucial for future surveys at the confusion limits of instruments like {\it Herschel} at 300-500\,$\mu$m and SCUBA2 at 850\,$\mu$m. 
Even as we wait for EVLA and the new generation of SKA precursors, our results show that the GMRT at 610\,MHz and even 325\,MHz
 can already currently provide sensitivity well matched to wide {\it Herschel} surveys.

It would be interesting to explore further whether the main properties which
characterize this sample, i.e. strong MIR emission, radio activity, and high stellar mass, are related. Some of these properties
are likely the result of biases introduced by our selection; however, this is
unlikely to be the case
 for all of them, especially for the radio properties. In
particular, a comparison between the starburst morphology (traced by young
stars, dust, PAHs or CO emission, as measured by ALMA or {\it JWST}) and the radio size would probe whether the radio
emission is produced by the starburst or by an AGN, and whether the parameters of the 
starburst are different from those of most classical SMGs and reveal a different star formation regime.

We have several multi-wavelength observations planned or in progress for
this sample to obtain better estimates of redshifts, dust temperatures, star formation rates, 
PAH luminosities, and AGN contributions, and thus constrain the dominant emission
processes, and investigate the evolution and clustering properties of these sources.

\begin{acknowledgements}

We thank Roy Kilgard, Jacqueline Bergeron, Attila Kov\'acs and Helmut Dannerbauer for their helpful contribution.
This work includes observations made with IRAM, which is supported by INSU/CNRS (France), MPG (Germany) and IGN (Spain).
Thanks to the staff of IRAM for their support and the anonymous observers of the MAMBO pool for observations presented here.
This work is based in part on observations made within the context of SWIRE survey with the {\it Spitzer Space Telescope}, 
which is operated by the Jet Propulsion Laboratory, California Institute of Technology under a contract with NASA.  
The VLA is operated by NRAO, 
the National Radio Astronomy Observatory, a facility of the National Science Foundation 
operated under cooperative agreement by Associated Universities, Inc.
We thank the staff of the GMRT who have made these observations 
possible. GMRT is run by the National Centre for Radio Astrophysics
of the Tata Institute of Fundamental Research.
The optical data come from KPNO (Kitt Peak National Observatory), National Optical Astronomy Observatory, 
which is operated by the Association of Universities for Research in Astronomy (AURA) under cooperative agreement with the National Science Foundation.
This work includes observations made with WFCAM/UKIRT. The UKIRT is operated by the Joint Astronomy Centre on behalf of the UK's Science and Technology Facilities Council. 
 MP acknowledges financial contribution from contract ASI-INAF I/016/07/0.

\end{acknowledgements}

\twocolumn
\bibliographystyle{aa.bst}

\bibliography{12742}

\clearpage

\begin{landscape}
\centering
\begin{table}[b]

\caption{\label{new} \label{NIR} Optical, Near-IR and {\it Spitzer} mid-IR data of the selected sample}
\begin{tabular}{llccccccccccccc}
\hline\hline{}
ID & IAU name &  $u$ & $g^{\prime}$ & $r^{\prime}$ & $i^{\prime}$ & $z$ & $J$ & $H$ & $K$ & $F_{\rm{3.6\,\mu m}}$ & $F_{\rm{4.5\,\mu m}}$ & $F_{\rm{5.8\,\mu m}}$ & $F_{\rm{8.0\,\mu m}}$ & $F_{\rm{24\,\mu m}}$ \\
\multicolumn{2}{c}{}   & (Vega) & (Vega) & (Vega) &  (Vega) & (Vega) & (Vega) & (Vega) & (Vega) & ($\mu$Jy) & ($\mu$Jy) & ($\mu$Jy) & ($\mu$Jy) & ($\mu$Jy) \\
\hline
L-1   & SWIRE3\_J104351.16+590057.9 &              25.3 &              25.6 &              25.0 &              24.3 &              23.5 &             22.1 &            21.1 & $\mathellipsis$ & 26.1$\pm$0.6 & 37.1$\pm$0.9 & 38.0$\pm$3.3 &  33.6$\pm$3.2 & 721.8$\pm$17.0 \\
L-2   & SWIRE3\_J104357.61+584921.5 &  \textgreater24.5 &  \textgreater24.9 &  \textgreater24.0 &  \textgreater23.2 &  \textgreater23.6 &             21.8 &            20.7 & $\mathellipsis$ & 27.6$\pm$0.7 & 34.8$\pm$1.0 & 42.9$\pm$3.8 & \textless40.0 & 625.0$\pm$17.3 \\
L-3   & SWIRE3\_J104409.98+584055.9 &              23.8 &              24.2 &              23.3 &              22.8 &              22.1 &             20.5 &            19.7 & $\mathellipsis$ & 41.2$\pm$1.2 & 45.4$\pm$0.9 & 47.7$\pm$5.7 &  38.8$\pm$3.1 & 579.2$\pm$16.5 \\
L-4   & SWIRE3\_J104425.97+584024.0 &  \textgreater24.5 &  \textgreater24.9 &  \textgreater24.0 &  \textgreater23.2 &  \textgreater23.6 &             21.7 &            20.3 & $\mathellipsis$ & 40.0$\pm$1.2 & 47.3$\pm$1.0 & 69.0$\pm$5.9 & \textless40.0 & 522.2$\pm$18.0 \\
L-5   & SWIRE3\_J104427.53+584309.8 &              23.4 &              23.9 &              23.4 &              22.9 &              22.0 &             20.4 &            19.6 & $\mathellipsis$ & 38.3$\pm$1.0 & 49.2$\pm$1.2 & 53.6$\pm$4.6 & \textless40.0 & 549.9$\pm$15.1 \\
L-6   & SWIRE3\_J104430.24+590701.6 &  \textgreater24.5 &  \textgreater24.9 &  \textgreater24.0 &  \textgreater23.2 &  \textgreater23.6 &             22.7 &            21.6 &            20.2 & 18.6$\pm$0.5 & 25.1$\pm$0.8 & 42.0$\pm$3.1 & \textless40.0 & 454.4$\pm$17.2 \\
L-7   & SWIRE3\_J104430.61+585518.4 &              25.6 &              25.6 &              24.6 &              23.8 &   $\mathellipsis$ &             21.1 &            20.3 &            19.2 & 41.3$\pm$0.8 & 49.5$\pm$1.1 & 55.3$\pm$4.0 &  53.9$\pm$3.8 & 870.0$\pm$17.2 \\
L-8   & SWIRE3\_J104439.98+591240.4 &  \textgreater24.5 &  \textgreater24.9 &  \textgreater24.0 &  \textgreater23.2 &  \textgreater23.6 &  $\mathellipsis$ & $\mathellipsis$ & $\mathellipsis$ & 21.0$\pm$0.6 & 33.1$\pm$0.9 & 45.9$\pm$3.3 & \textless40.0 & 496.9$\pm$17.1 \\
L-9   & SWIRE3\_J104440.25+585928.4 &  \textgreater24.5 &  \textgreater24.9 &  \textgreater24.0 &  \textgreater23.2 & \textgreater23.6 &    23.9 &            22.2 &            21.6 & 28.2$\pm$0.7 & 39.2$\pm$1.0 & 44.8$\pm$3.9 & \textless40.0 & 674.2$\pm$18.7 \\
L-10  & SWIRE3\_J104549.75+591903.5 &  \textgreater24.5 &  \textgreater24.9 &  \textgreater24.0 &  \textgreater23.2 &              23.4 &             22.7 &            20.9 &            19.5 & 31.3$\pm$0.5 & 45.3$\pm$0.8 & 61.4$\pm$3.2 &  60.3$\pm$3.4 & 635.2$\pm$16.6 \\
L-11  & SWIRE3\_J104556.90+585318.9 &              23.8 &              24.6 &              24.2 &              23.4 &  \textgreater23.6 &             21.8 &            20.6 &            19.9 & 26.4$\pm$0.7 & 33.6$\pm$1.0 & 39.5$\pm$3.6 &  36.9$\pm$3.6 & 650.3$\pm$17.5 \\
L-12  & SWIRE3\_J104601.79+590916.8 &  \textgreater24.5 &              25.1 &              23.7 &              23.2 &              22.4 &  $\mathellipsis$ & $\mathellipsis$ & $\mathellipsis$ & 72.1$\pm$1.0 & 91.4$\pm$1.5 & 93.5$\pm$3.9 &  74.0$\pm$3.9 & 976.0$\pm$16.3 \\
L-13  & SWIRE3\_J104602.09+592003.1 &  \textgreater24.5 &  \textgreater24.9 &  \textgreater24.0 &  \textgreater23.2 &  \textgreater23.6 &             22.8 &            21.5 &            20.8 & 15.9$\pm$0.4 & 22.1$\pm$0.7 & 32.7$\pm$3.1 &  29.2$\pm$3.2 & 415.2$\pm$16.3 \\
L-14  & SWIRE3\_J104638.67+585612.6 &              25.9 &              25.8 &              24.9 &              25.3 &  \textgreater23.6 &             23.4 &            21.2 &            19.5 & 29.5$\pm$0.7 & 38.8$\pm$1.0 & 58.2$\pm$3.7 &  42.1$\pm$3.5 & 610.9$\pm$17.0 \\
L-15  & SWIRE3\_J104656.47+590235.5 &              25.5 &              25.3 &              24.7 &              23.8 &              22.8 &             20.7 &            19.7 &            18.7 & 57.6$\pm$0.8 & 72.6$\pm$1.2 & 76.3$\pm$3.5 &  38.2$\pm$3.1 & 419.4$\pm$17.6 \\
L-16  & SWIRE3\_J104702.82+591836.9 &  \textgreater24.5 &  \textgreater24.9 &  \textgreater24.0 &  \textgreater23.2 &              23.7 &             22.0 &            20.8 &            19.8 & 30.0$\pm$0.6 & 37.2$\pm$1.0 & 44.1$\pm$3.5 & \textless40.0 & 513.8$\pm$18.7 \\
L-17  & SWIRE3\_J104704.97+592332.3 &              23.6 &              24.3 &              23.6 &              23.1 &              22.3 &             20.7 &            20.0 &            19.0 & 34.7$\pm$0.5 & 50.5$\pm$1.3 & 59.4$\pm$2.9 & \textless40.0 & 646.6$\pm$17.9 \\
L-18  & SWIRE3\_J104706.82+585848.2 &              25.3 &              25.3 &              24.8 &              24.0 &              22.8 &             20.6 &            19.7 &            18.5 & 57.1$\pm$0.9 & 72.4$\pm$1.2 & 77.0$\pm$3.5 &  41.6$\pm$3.3 & 593.4$\pm$19.2 \\
L-19  & SWIRE3\_J104708.79+590627.2 &              25.5 &              26.9 &              26.4 &              24.4 &              23.3 &             21.1 &            20.3 &            19.2 & 35.9$\pm$0.8 & 50.0$\pm$1.1 & 77.4$\pm$4.0 &  52.1$\pm$3.8 & 720.8$\pm$19.2 \\
L-20  & SWIRE3\_J104717.96+590231.8 &              25.6 &              25.3 &              24.5 &              24.4 &              23.6 &             21.5 &            20.4 &            19.2 & 54.8$\pm$0.9 & 67.1$\pm$1.3 & 71.2$\pm$3.6 & \textless40.0 & 617.0$\pm$19.1 \\
L-21  & SWIRE3\_J104718.63+584318.1 &  \textgreater24.5 &  \textgreater24.9 &  \textgreater24.0 &  \textgreater23.2 &   $\mathellipsis$ &             21.6 &            20.5 & $\mathellipsis$ & 34.4$\pm$0.6 & 45.0$\pm$0.9 & 54.5$\pm$2.9 &  46.1$\pm$3.6 & 447.0$\pm$16.1 \\
L-22  & SWIRE3\_J104720.49+591043.6 &              25.6 &              25.2 &              24.2 &              23.7 &  \textgreater23.6 &             21.5 &            20.7 &            19.5 & 31.6$\pm$0.6 & 41.6$\pm$0.8 & 55.6$\pm$3.6 &  54.5$\pm$3.4 & 435.1$\pm$15.9 \\
L-23  & SWIRE3\_J104726.44+585213.3 &  \textgreater24.5 &  \textgreater24.9 &  \textgreater24.0 &  \textgreater23.2 &  \textgreater23.6 &             22.5 &            21.4 &            20.7 & 18.9$\pm$0.6 & 27.2$\pm$0.9 & 37.5$\pm$3.5 &  37.4$\pm$3.3 & 434.0$\pm$17.1 \\
L-24  & SWIRE3\_J104732.10+585127.1 &              24.2 &              24.5 &              24.2 &              23.5 &              22.6 &             21.4 &            20.2 &            19.4 & 28.3$\pm$0.7 & 34.3$\pm$0.9 & 37.7$\pm$3.3 &  34.0$\pm$3.3 & 419.7$\pm$19.3 \\
L-25  & SWIRE3\_J104738.32+591010.0 &              24.1 &              24.8 &              23.9 &              23.2 &  \textgreater23.6 &             20.6 &            19.7 &            19.0 & 44.4$\pm$0.6 & 51.0$\pm$0.8 & 52.3$\pm$3.1 &  47.7$\pm$3.3 & 722.6$\pm$17.4 \\
L-26  & SWIRE3\_J104744.09+590624.8 &  \textgreater24.5 &  \textgreater24.9 &  \textgreater24.0 &  \textgreater23.2 &              23.5 &             21.3 &            20.2 &            19.3 & 36.3$\pm$0.5 & 43.5$\pm$0.8 & 46.8$\pm$2.9 &  36.9$\pm$3.3 & 494.4$\pm$18.5 \\
L-27  & SWIRE3\_J104744.59+591413.4 &              24.6 &              24.5 &              23.9 &              23.4 &              22.6 &             21.4 &            20.3 &            19.6 & 27.8$\pm$0.5 & 34.6$\pm$0.7 & 41.8$\pm$3.4 & \textless40.0 & 522.9$\pm$17.9 \\
L-28  & SWIRE3\_J104757.59+590451.7 &              25.7 &              25.9 &              25.1 &              24.4 &              23.2 &             21.1 &            20.3 &            19.3 & 32.6$\pm$0.6 & 40.7$\pm$0.8 & 48.5$\pm$3.2 & \textless40.0 & 418.7$\pm$18.8 \\
L-29  & SWIRE3\_J104807.86+591119.5 &              24.6 &              24.8 &              24.2 &              23.5 &              22.5 &             20.8 &            19.8 &            19.1 & 40.7$\pm$0.6 & 49.3$\pm$0.8 & 64.2$\pm$3.3 &  38.6$\pm$3.2 & 407.2$\pm$19.1 \\
L-30  & SWIRE3\_J104816.81+584658.5 &              24.1 &              24.7 &              24.3 &              23.6 &  \textgreater23.6 &  $\mathellipsis$ & $\mathellipsis$ & $\mathellipsis$ & 35.6$\pm$0.6 & 45.7$\pm$0.9 & 49.1$\pm$3.3 & \textless40.0 & 469.9$\pm$16.4 \\
L-31  & SWIRE3\_J104830.71+585659.3 &  \textgreater24.5 &  \textgreater24.9 &  \textgreater24.0 &  \textgreater23.2 &   $\mathellipsis$ & \textgreater24.2 &            22.8 &            21.0 & 17.4$\pm$0.5 & 24.2$\pm$0.9 & 43.8$\pm$3.5 & \textless40.0 & 639.9$\pm$15.6 \\
L-32  & SWIRE3\_J104841.00+590056.8 &  \textgreater24.5 &  \textgreater24.9 &  \textgreater24.0 &  \textgreater23.2 &   $\mathellipsis$ &             22.9 &            20.9 &            20.0 & 27.0$\pm$0.6 & 37.1$\pm$0.7 & 44.2$\pm$3.3 &  30.6$\pm$3.0 & 406.2$\pm$19.3 \\
L-33  & SWIRE3\_J104848.23+585059.3 &              20.1 &              25.1 &              24.3 &              23.8 &              22.9 &             21.4 &            20.3 &            19.7 & 32.1$\pm$0.7 & 37.6$\pm$0.9 & 51.4$\pm$3.7 &  33.4$\pm$3.8 & 559.7$\pm$20.3 \\

\hline

\end{tabular}
\begin{flushleft}
The optical magnitude limits correspond to 90\% completeness values derived in the field where our sample was selected, but note that the coverage of the optical observations is not uniform across the field.\\
The optical data come from NOAO and the Near-IR magnitudes from UKIRT (see text)
\end{flushleft}
\end{table}
\end{landscape}

\clearpage
\onecolumn

\begin{table}[!t]
\centering
\caption{Related \textit{Spitzer}-selected $z\sim2$ ULIRGs samples.}
\label{samples}

\begin{tabular}{lcccccc}
\hline\hline
Sample         & N & $F_{\rm{24\mu m}}$ & $< F_{\rm{24\,\mu m}}>$ & IR color selection & $r$    & density    \\
               &   & $\mu$Jy  & $\mu$Jy &      & Vega & deg$^{-2}$ \\
This work                      &     33   & \textgreater400 & 566  & $F_{\rm{3.6\,\mu m}}$\textless $F_{\rm{4.5\,\mu m}}$\textless $F_{\rm{5.8\,\mu m}}$\textgreater $F_{\rm{8.0\,\mu m}}$   & \textgreater23  & 65 \\
Lonsdale et al.\ 2008                           &     61  & \textgreater400 & 819  &  $F_{\rm{3.6\,\mu m}}$\textless $F_{\rm{4.5\,\mu m}}$\textless $F_{\rm{5.8\,\mu m}}$\textgreater $F_{\rm{8.0\,\mu m}}$  & \textgreater23    & 55  \\
Farrah et al.\ 2008            & 32  & \textgreater500  & 726 & $F_{\rm{3.6\,\mu m}}$\textless $F_{\rm{4.5\,\mu m}}$\textgreater $F_{\rm{5.8\,\mu m}}$ and $F_{\rm{4.5\,\mu m}}$\textgreater $F_{\rm{8.0\,\mu m}}$    & \textgreater23    &  34  \\
Huang et al.\ 2009$^{a}$             &      12  & \textgreater500 & 689 &                                                       0\textless[3.6]-[4.5]\textless0.4   &                   &   \\
                               &          & &                 &                                                    -0.7\textless[3.6]-[8.0]\textless0.5   &                   \\
Magliocchetti et al.\ 2007     &     793  & \textgreater350  & &                                                                                          & \textgreater25.5  & 200  \\
Yan et al.\ 2005               &      52  & \textgreater900  & &  $\log_{10}(\nu f_{\nu}(\rm{24\mu m})/\nu {\textit f}_{\nu}(\rm{8\mu m}))$\textgreater0.5      &                   &    \\
                               &          &                  & & $\log_{10}(\nu f_{\nu}(\rm{24\mu m})/\nu {\textit f}_{\nu}(\rm{0.7\mu m})$)\textgreater1.0  &                   &      \\\hline
\end{tabular}
\onecolumn
\begin{flushleft}
$^{a}$ The sample of \citet{Huan08} differs by only one source from the one of \citet{Youn09}.
\end{flushleft}
\end{table}

\begin{table}
\caption{\label{stack2} Derived values from the stacked flux densities$^a$.}
\centering
\begin{tabular}{lcccccccccccc}

\hline\hline
Sample & N & z & $F_{\rm{1.2\,mm}}$ & $F_{\rm{20\,cm}}$ & $F_{\rm{1.2\,mm}}$/$F_{\rm{24\,\mu m}}$ & $L_{\rm{1.4\,GHz}}$ & $L_{\rm{FIR,mm}}$$^{c}$ & $L_{\rm{FIR,radio}}$ & SFR$_{\rm{mm}}$ & SFR$_{\rm{radio}}$ & q \\
       &   &   & (mJy)         & ($\mu$Jy)    & & (10$^{24}$\,W\,Hz$^{-1}$) &  \multicolumn{2}{c}{(10$^{12}$\,$L_{\odot}$)} &  \multicolumn{2}{c}{($M_{\odot}$\,yr$^{-1}$)} & \\
\hline
All    & 33 & 2.08 & 1.56$\pm$0.22 & 92.6$\pm$12.0 & 2.76$\pm$0.50 & 2.03$\pm$0.26 & 2.45$\pm$0.35 & 4.14$\pm$0.53 & 441 & 745 & 2.11$\pm$0.12 \\
S/N$>$3$^{b}$ & 13 & 2.11 & 2.83$\pm$0.14 & 128.5$\pm$26.0 & 4.94$\pm$0.51 & 2.89$\pm$0.58 & 3.99$\pm$0.20 & 5.90$\pm$1.18 & 718 & 1062 & 2.17$\pm$0.11\\
S/N$<$3$^{b}$ & 20 & 2.06 & 0.72$\pm$0.18 & 69.3$\pm$7.4 & 1.28$\pm$0.40 & 1.48$\pm$0.16 & 1.38$\pm$0.35 & 3.02$\pm$0.32 & 248 & 543 & 2.00$\pm$0.16\\
\hline
\end{tabular}
\begin{flushleft}
$^a$The average stacked flux densities are computed with equal weights to avoid biases. Their approximate rms are computed as the standard deviation of the mean.\\
$^{b}$S/N at 1.2\,mm.\\
$^{c}$$L_{\rm{FIR,mm}}$ is computed with (Eq. 3) and the dust temperatures coming from the stacked SED (see text): 37\,K for the entire sample, 36\,K for the 13 sources with S/N\,$>$\,3 at 1.2\,mm, and 39\,K for the 20 sources with S/N\,$<$\,3.\\
\end{flushleft}
\end{table}

\clearpage
\centering
\begin{table}[b]
\caption{\label{res}1.2\,mm and radio data.}
\begin{tabular}{lcccccccc}
\hline\hline

Source ID  & $F_{\rm{1.2\,mm}}$$^a$ & $F_{\rm{20\,cm}}$$^b$ & $F_{\rm{50\,cm}}$$^b$ & $F_{\rm{90\,cm}}$$^b$ & major axis & minor axis &  $\alpha$$^c$ & q \\
  & (mJy) & ($\mu$Jy) & ($\mu$Jy) & ($\mu$Jy) & ($\arcsec$) & ($\arcsec$) &  &  \\
\hline
\multicolumn{9}{c}{1.2\,mm detections S/N$>$4}\\
\hline
L-1  	&	  2.95$\pm$0.66 	&		   76.3$\pm$9.0 	&	 142$\pm$19 	&	      327$\pm$74 	&	  \textless2.9 	&	   X 	&	 $-$0.84$\pm$0.18		&	 2.49$^{+0.20}_{-0.20}$ \\																	
\vspace{0.4 ex}	
L-9  	&	  4.00$\pm$0.55 	&		  116.5$\pm$9.2 	&	 195$\pm$22 	&	      266$\pm$69 	&	           1.9	&	 0.7	&	 $-$0.54$\pm$0.03		&	 2.35$^{+0.20}_{-0.20}$ \\
\vspace{0.4 ex}																		
L-11 	&	  3.08$\pm$0.58 	&		  314.8$\pm$4.1 	&	 425$\pm$43 	&	      651$\pm$72 	&	             X 	&	   X 	&	 $-$0.45$\pm$0.06		&	 2.06$^{+0.25}_{-0.25}$ \\
\vspace{0.4 ex}																		
L-22 	&	  3.41$\pm$0.73 	&		  100.9$\pm$7.1 	&	 238$\pm$26 	&	      447$\pm$70 	&	  \textless1.7 	&	   X 	&	 $-$1.02$\pm$0.01		&	 2.00$^{+0.23}_{-0.23}$ \\
\hline
\multicolumn{9}{c}{1.2\, mm tentative detections S/N$>$3}\\
\hline
L-5  	&	  2.75$\pm$0.76 	&		  71.0$\pm$20.0 	&	 177$\pm$23 	&	      322$\pm$74 	&	  \textless4.5 	&	   X 	&	 $-$1.03$\pm$0.06		&	 2.28$^{+0.21}_{-0.21}$ \\
\vspace{0.4 ex}																		
L-14 	&	  2.13$\pm$0.71 	&		  159.5$\pm$6.0 	&	 322$\pm$33 	&	      426$\pm$70 	&	           0.7	&	   X 	&	 $-$0.73$\pm$0.09		&	 1.81$^{+0.26}_{-0.26}$ \\
\vspace{0.4 ex}																		
L-15 	&	  2.36$\pm$0.62 	&	           68.6$\pm$3.7 	&	 181$\pm$21 	&	      258$\pm$69 	&	  \textless1.0 	&	   X 	&	 $-$1.03$\pm$0.14		&	 2.31$^{+0.20}_{-0.20}$ \\
\vspace{0.4 ex}																		
L-17 	&	  2.24$\pm$0.64 	&		 341.0$\pm$33.0 	&	 756$\pm$78 	&	     1052$\pm$75 	&	             X 	&	   X 	&	 $-$0.76$\pm$0.10		&	 1.68$^{+0.20}_{-0.20}$ \\
\vspace{0.4 ex}																		
L-20 	&	  2.66$\pm$0.78 	&		   51.1$\pm$4.7 	&	 166$\pm$19 	&	      264$\pm$72 	&	  \textless1.5 	&	   X 	&	 $-$1.25$\pm$0.18		&	 1.95$^{+0.17}_{-0.17}$ \\
\vspace{0.4 ex}																		
L-21 	&	  3.09$\pm$0.81 	&		 138.3$\pm$35.8 	&	 176$\pm$38 	&	      350$\pm$76 	&	           4.1	&	 1.8	&	 $-$0.64$\pm$0.28		&	 2.51$^{+0.37}_{-0.37}$ \\
\vspace{0.4 ex}																		
L-23 	&	  3.13$\pm$0.86 	&		   86.4$\pm$7.2 	&	 172$\pm$21 	&	      242$\pm$71 	&	  \textless1.3 	&	   X 	&	 $-$0.77$\pm$0.07		&	 2.23$^{+0.23}_{-0.23}$ \\
\vspace{0.4 ex}																		
L-25 	&	  2.56$\pm$0.74 	&		   69.0$\pm$9.0 	&	 166$\pm$20 	&	      348$\pm$72 	&	  \textless1.4 	&	   X 	&	 $-$1.09$\pm$0.04		&	 2.63$^{+0.25}_{-0.25}$ \\
\vspace{0.4 ex}																		
L-27 	&	  2.48$\pm$0.73 	&		  77.2$\pm$13.7 	&	 106$\pm$18 	&	    \textless219 	&	  \textless2.1 	&	   X 	&	 $-$0.38$\pm$0.39		&	 2.47$^{+0.27}_{-0.27}$ \\
\hline
\multicolumn{9}{c}{Other non detections at 1.2\,mm}\\
\hline																		
L-2  	&	  \textless3.04 (1.46$\pm$0.79) 	&		  \textless42.0 	&	  90$\pm$17 	&	    \textless215 	&	        X 	&	   X 	&      [$-$0.64]		&	      \\
\vspace{0.4 ex}																		
L-3  	&	  \textless3.33 (1.04$\pm$1.11) 	&		 105.0$\pm$35.0 	&	  50$\pm$17 	&	    \textless227 	&  \textless7.6 	&	   X 	& +0.89$\pm$0.85		&	      \\
\vspace{0.4 ex}																		
L-4  	&	  \textless3.42 (0.76$\pm$1.14) 	&		  \textless93.0 	&	  76$\pm$17 	&	    \textless220 	&	          X 	&	   X 	&	[$-$0.64]		&	      \\
\vspace{0.4 ex}																		
L-6  	&	  \textless2.22 (0.59$\pm$0.74) 	&		   37.1$\pm$6.4 	&	  69$\pm$12 	&	    \textless202 	&	  \textless1.4 	&	   X 	& $-$0.75$\pm$0.42		&	  \\
\vspace{0.4 ex}																		
L-7  	&	  \textless2.97 (1.39$\pm$0.79) 	&		 101.7$\pm$13.3 	&	 200$\pm$23 	&	    \textless216 	&	           2.0	&	 0.8	& $-$0.81$\pm$0.29		&	  \\
\vspace{0.4 ex}																		
L-8  	&	  \textless2.70 (-0.36$\pm$0.90) 	&		 124.0$\pm$14.0 	&	 219$\pm$25 	&	      250$\pm$72 	&	           1.4	&	   X 	& $-$0.58$\pm$0.13		&	           \\
\vspace{0.4 ex}																		
L-10 	&	  \textless3.01 (1.39$\pm$0.81) 	&		  67.2$\pm$11.7 	&	  94$\pm$17 	&	 $\mathellipsis$ 	&	  \textless3.7 	&	   X 	& $-$0.40$\pm$0.37		&	 \\
\vspace{0.4 ex}																		
L-12 	&	  \textless3.64 (2.06$\pm$0.79) 	&		   94.4$\pm$4.7 	&	 182$\pm$28 	&	      315$\pm$69 	&	  \textless1.3 	&	   X 	& $-$0.81$\pm$0.02		&	  \\
\vspace{0.4 ex}																		
L-13 	&	  \textless3.15 (-0.49$\pm$1.05) 	&		  \textless39.0 	&	  58$\pm$16 	&	    \textless219 	&	             X 	&	   X 	&        [$-$0.64]		&	          \\
\vspace{0.4 ex}																		
L-16 	&	  \textless2.58 (0.51$\pm$0.86) 	&	        $\mathellipsis$ 	&	 145$\pm$32 	&	      319$\pm$77 	&	             X 	&	   X 	& $-$1.25$\pm$0.76		&	       \\
\vspace{0.4 ex}																		
L-18 	&	  \textless2.43 (-0.72$\pm$0.81) 	&		   93.2$\pm$9.4 	&	 162$\pm$19 	&	    \textless209 	&	           2.4	&	   X 	& $-$0.67$\pm$0.36		&	           \\
\vspace{0.4 ex}																		
L-19 	&	  \textless2.91 (0.89$\pm$0.97) 	&		 102.2$\pm$12.0 	&	 138$\pm$18 	&	      231$\pm$69 	&	           3.9	&	 1.8	& $-$0.44$\pm$0.13		&	  \\
\vspace{0.4 ex}																		
L-24 	&	  \textless2.64 (0.14$\pm$0.88) 	&		  \textless24.0 	&	  45$\pm$14 	&	    \textless211 	&	             X 	&	   X 	&        [$-$0.64]		&	        \\
\vspace{0.4 ex}																		
L-26 	&	  \textless2.58 (0.37$\pm$0.86) 	&		  54.9$\pm$13.4 	&	  84$\pm$14 	&	      255$\pm$72 	&	           3.5	&	   X 	& $-$0.86$\pm$0.53		&	  \\
\vspace{0.4 ex}																		
L-28 	&	  \textless2.96 (1.64$\pm$0.66) 	&		  \textless23.0 	&	  47$\pm$12 	&	    \textless213 	&	             X 	&	   X 	&        [$-$0.64]		&	      \\
\vspace{0.4 ex}																		
L-29 	&	  \textless2.94 (0.45$\pm$0.98) 	&		  55.0$\pm$12.0 	&	  76$\pm$16 	&	    \textless216 	&	             X 	&	   X 	& $-$0.39$\pm$0.52		&	 \\
\vspace{0.4 ex}																		
L-30 	&	  \textless1.89 (0.42$\pm$0.63) 	&		  \textless70.0 	&	  84$\pm$18 	&	    \textless236 	&	             X 	&	   X 	&        [$-$0.64]		&	  \\
\vspace{0.4 ex}																	
L-31 	&	  \textless2.82 (1.62$\pm$0.60) 	&		 121.0$\pm$39.0 	&	 157$\pm$20 	&	    \textless221 	&	           6.6	&	   X 	& $-$0.31$\pm$0.50		&	  \\
\vspace{0.4 ex}																		
L-32 	&	  \textless2.89 (1.61$\pm$0.64) 	&		  \textless48.0 	&	  60$\pm$16 	&	 $\mathellipsis$ 	&	             X 	&	   X 	&        [$-$0.64]		&	        \\
\vspace{0.4 ex}																		
L-33 	&	  \textless3.27 (-0.28$\pm$1.09) 	&		 109.0$\pm$33.0 	&	  87$\pm$19 	&	    \textless235 	&	 \textless10.0 	&	   X 	&   +0.27$\pm$0.61		&	         \\
\hline
\end{tabular}
\begin{flushleft}
$^a$The limits on $F_{\rm{1.2\,mm}}$  are '3$\sigma$' limits, except for the sources with 1.5\,$<\,$S/N$\,<$\,3 where they are the observed value plus 2\,$\sigma$. 
  The values in brackets are the observed values used in stacks.\\
$^b$The radio flux density limits are '3$\sigma$' limits.\\
$^c$The values of $\alpha$ in brackets show the cases where $\alpha$ cannot be determined and is assumed equal to -0.64  (see text).\\

\end{flushleft}
\end{table}

\clearpage
\onecolumn
\begin{table}
\caption{\label{lum} Luminosities and star formation rates.}
\centering
\begin{tabular}{lccccccccl}

\hline\hline
Source & $z_{phot}$ & $L_{\rm{1.4\,GHz}}$ $^{b}$           & $L_{\rm{FIR,mm}}$         & $L_{\rm{FIR,radio}}$ $^{c}$ &      $SFR_{\rm{mm}}$ $^{d}$&      $SFR_{\rm{radio}}$ $^{d}$ & $\nu$$L_{\nu}$(1.6\,$\mu$m)    & $M_{\star}$                                  & SMG/SFRG \\
       &            & (10$^{24}$\,W\,Hz$^{-1}$) &  \multicolumn{2}{c}{(10$^{12}$\,$L_{\odot}$)} &  \multicolumn{2}{c}{($M_{\odot}$\,yr$^{-1}$)} & (10$^{11}$\,$L_{\odot}$) & \multicolumn{1}{c}{(10$^{11}$\,$M_{\odot}$)} & \\
\hline
\vspace{0.4 ex}																	
L-1  &       1.92 &            1.60 & 4.63$\pm$1.04 &    3.27$\pm$0.81 &                      834 & 588 & 1.67 &   1.03 & SMG        \\	
\vspace{0.4 ex}																	
L-2  &       1.92 &            0.96 & \textless4.77 &    1.96$\pm$0.39 &             \textless859 & 353 & 1.61 &   1.08 & weak Radio? \\	
\vspace{0.4 ex}														 			
L-3  &       1.73 &            0.33 & \textless5.22 &  0.66$\pm$0.52 &  \textless941 & \textless284 & 1.66 & 1.13 & weak Radio \\	
\vspace{0.4 ex}																	
L-4  &       2.13 &            1.02 & \textless5.37 &    2.08$\pm$3.08 &    \textless966 & \textless1664 & 3.02 &   2.85 & SFRG?      \\	
\vspace{0.4 ex}																	
L-5  &       2.04 &            2.40 & 4.32$\pm$1.19 &    4.90$\pm$0.92 &                      777 & 883 & 2.53 &   1.36 & SMG?        \\	
\vspace{0.4 ex}																	
L-6  &       2.26 &            1.10 & \textless3.49 &  2.25$\pm$0.79 &           \textless627 & 405 & 2.07 & 1.07 & SFRG? \\	
\vspace{0.4 ex}																	
L-7  &       1.89 &            2.15 & \textless4.66 &  4.35$\pm$1.06 &                    \textless839 & 788 & 2.19 & 1.55 & ?      \\	
\vspace{0.4 ex}																	
L-8  &       2.42 &            3.80 & \textless4.24 &    [7.75$\pm$1.09] &             \textless763 & [1396] & 2.89 &   [1.52] & SFRG       \\	
\vspace{0.4 ex}																	
L-9  &       2.33 &            2.99 & 6.28$\pm$0.86 &    6.11$\pm$1.94 &                     1130 & 1100 & 2.71 &   2.66 & SMG        \\	
\vspace{0.4 ex}																	
L-10 &       2.45 &            1.56 & \textless4.73 &    3.19$\pm$1.05 &                      \textless851 & 574 & 4.01 &  2.31 & ?      \\	
\vspace{0.4 ex}																	
L-11 &       1.92 &            4.52 & 4.84$\pm$0.91 &  [9.21$\pm$3.63] &                   870 & [1658] & 1.57 & [0.78] & SMG        \\	
\vspace{0.4 ex}																	
L-12 &       1.71 &            1.56 & \textless5.71 &    3.19$\pm$1.12 &                    \textless1029 & 575 & 3.19 &   3.09 & ?       \\ 	
\vspace{0.4 ex}																	
L-13 &       2.22 &            0.86 & \textless4.95 &    1.75$\pm$0.89 &    \textless891 & 315 & 1.60 &   0.87 & weak Radio \\	
\vspace{0.4 ex}																	
L-14 &       2.36 &            5.53 & 3.34$\pm$1.11 &  [11.30$\pm$2.99] &                   602 & [2029] & 3.37 & [1.74] & SMG?        \\	
\vspace{0.4 ex}								 							
L-15 &       1.86 &            1.92 & 3.71$\pm$0.97 &    3.93$\pm$0.76 &                      667 & 707 & 3.09 &   3.21 & SMG?        \\	
\vspace{0.4 ex}								 									
L-16 &       1.73 &            1.36 & \textless4.05 & 2.77$\pm$1.51  & \textless729 & 499 & 1.34 &   0.92 & SFRG?       \\	
\vspace{0.4 ex}																	
L-17 &       1.90 &            7.90 & 3.52$\pm$1.00 & [16.12$\pm$2.70] &                   633 & [2902] & 2.27 & [1.07] & SMG?        \\	
\vspace{0.4 ex}																
L-18 &       1.85 &            1.60 & \textless3.82 &  3.27$\pm$0.86 &           \textless688 & 589 & 3.05 & 2.17 & SFRG       \\	
\vspace{0.4 ex}																	
L-19 &       1.69 &            1.08 & \textless4.57 &  2.21$\pm$0.53 &           \textless822 & 398 & 1.67 & 0.99 & ?      \\	
\vspace{0.4 ex}																	
L-20 &       2.69 &            4.98 & 4.18$\pm$1.22 &   10.20$\pm$0.95 &                      752 & 1830 & 5.68 &   7.16 & SMG?        \\	
\vspace{0.4 ex}																	
L-21 &       1.78 &            1.61 & 4.85$\pm$1.27 &    3.23$\pm$1.94 &                     873 & 592 & 1.74 &   1.12 & SMG?        \\	
\vspace{0.4 ex}																	
L-22 &       2.57 &            5.70 & 5.35$\pm$1.15 &   11.62$\pm$4.323 &                     964 & 2092 & 4.11 &   3.06 & SMG        \\	
\vspace{0.4 ex}																	
L-23 &       2.38 &            3.08 & 4.91$\pm$1.35 &    6.28$\pm$1.63 &                      885 & 1130 & 2.27 &   1.27 & SMG?        \\	
\vspace{0.4 ex}																
L-24 &       2.07 &            0.57 & \textless4.14 &    1.16$\pm$0.29 &    \textless746 & \textless209 & 1.82 &   1.02 & weak Radio \\	
\vspace{0.4 ex}																	
L-25 & 1.47$^{a}$ &            1.01 & 4.02$\pm$1.16 &    2.07$\pm$0.59 &                      723 & 372 & 1.29 &   0.84 & SMG?        \\	
\vspace{0.4 ex}														 			
L-26 &       1.72 &            0.73 & \textless4.05 &    1.49$\pm$1.07 & \textless729 & \textless579    & 1.55 & 1.57 & weak Radio      \\	
\vspace{0.4 ex}																	
L-27 &       2.20 &            1.41 & 3.89$\pm$1.15 &    2.87$\pm$0.96 &                      701 & 517 & 2.18 &   1.21 & SMG?        \\	
\vspace{0.4 ex}																	
L-28 &       1.88 &            0.48 & \textless4.65 &    0.98$\pm$0.20 &             \textless836 & \textless176 & 1.79 &   1.29 & ?       \\	
\vspace{0.4 ex}																	
L-29 &       1.93 &            0.77 & \textless4.62 &    1.58$\pm$0.68 &   \textless831 & 284 & 2.33 &   1.62 & weak Radio      \\	
\vspace{0.4 ex}																	
L-30 &       2.38 &            1.45 & \textless2.97 &    2.96$\pm$2.39 &    \textless534 & \textless1290 & 3.17 &   2.01 & SFRG      \\	
\vspace{0.4 ex}																	
L-31 &       2.74 &            3.13 & \textless4.43 &  6.39$\pm$3.43 &          \textless797 & 1150 & 3.65 & 1.37 & ?       \\	
\vspace{0.4 ex}																	
L-32 &       2.38 &            1.03 & \textless4.53$\pm$1.00 &   2.10$\pm$1.54 &            \textless815 & \textless278 & 2.77 &   1.84 & ?       \\	
\vspace{0.4 ex}																	
L-33 &       2.18 &            0.92 & \textless5.13 &  1.87$\pm$0.97 & \textless924 & 337 & 2.47 & 1.49 & weak Radio?      \\	
\hline																	

\end{tabular}
\onecolumn
\begin{flushleft}
$^a$Spectroscopic redshift from \citet{Bert07a}\\
$^b$The value of $L_{\rm{1.4\,GHz}}$, the rest-frame luminosity at 1.4\,GHz, is the weighted average of the two determinations of $L_{\rm{1.4\,GHz}}$ from 20\,cm and 50\,cm flux densities when both are available.\\
$^c$The values of $L_{\rm{FIR,radio}}$, the infrared SFR derived from the radio flux densities and the stellar mass are dubious for sources with high 20\,cm/1.2\,mm flux densities ratios. They are reported, in brackets.\\
$^d$SFR calculated by Kennicutt's formula (Eq. 6) with a Salpeter IMF (see text).\\
The upper limits of $L_{\rm{FIR,mm}}$ and SFR$_{\rm{mm}}$ are '3$\sigma$' limits, except for the sources with 1.5\,$<\,$S/N$\,<$\,3 where they are the observed value plus 2\,$\sigma$.\\ 
The upper limits of $L_{\rm{FIR,radio}}$ and SFR$_{\rm{radio}}$ are '3$\sigma$' limits.\\
\end{flushleft}
\end{table}

\clearpage
\onecolumn

\begin{table}
\caption{\label{stack} Results from the co-added MIPS images.}
\centering
\begin{tabular}{lccccccc}

\hline\hline
Sample & N & $<$ $F_{\rm{24\,\mu m}}$ $>$ & $F_{\rm{70\,\mu m}}$ & $F_{\rm{160\,\mu m}}$ & $< F_{\rm{1.2\,mm}} >$ & $F_{\rm{160\,\mu m}}$/$< F_{\rm{1.2\,mm}}>$$^a$ & $F_{\rm{70\,\mu m}}$/ $< F_{\rm{24\,\mu m}} >$$^a$\\
       &   &  ($\mu$Jy)&\multicolumn{2}{c}{from stack median (mJy)} & (mJy) & & \\
\hline
\multicolumn{8}{c}{GO 30391 data}\\
\hline
{\bf All}                   & {\bf21} &    {\bf571$\pm$27} & {\bf2.78$\pm$0.33 } & {\bf13.47$\pm$2.76} & {\bf1.67$\pm$0.27}& {\bf8.1$\pm$3.0} &   {\bf4.9$\pm$0.8}	\\
1.2\,mm SNR$>$3       & 10 &   585$\pm$32 & 2.56$\pm$0.38 & 14.63$\pm$4.05 &  2.77$\pm$0.18 &  5.3$\pm$1.8 & 4.4$\pm$0.9 \\
1.2\,mm SNR$<$3       & 11 &   559$\pm$44 & 3.08$\pm$0.67 & 11.95$\pm$4.08 & 0.68$\pm$0.24 & 17.6$\pm$12.2 & 5.5$\pm$1.6 \\
\hline
\multicolumn{8}{c}{\citet{Youn09} data}\\
\hline
All		      & 12 & 	680$\pm$50 & 3.3$\pm$0.6  & 21.7$\pm$7.9 & 1.4$\pm$0.3 & 15.5$\pm$9.0 & 4.9$\pm$1.3 \\
\hline
\end{tabular}
\begin{flushleft}
$^{a}$$F_{\rm{160\,\mu m}}$/$< F_{\rm{1.2\,mm}} >$ and $F_{\rm{70\,\mu m}}$/ $< F_{\rm{24\,\mu m}} >$ are computed with the stack medians for $F_{\rm{70\,\mu m}}$ and $F_{\rm{160\,\mu m}}$
.\\
\end{flushleft}
\onecolumn
\end{table}

\Online
\newpage
\onecolumn
\begin{flushleft}
\begin{appendix}

\section{SED fits and photometric redshifts}\label{app}

The optical ($Ugriz$), NIR (JHK), and MIR (3.6--24\,$\mu$m) SEDs of each
source have been fitted with a library of 18 star-forming galaxy
templates~\citep{Poll07} using the {\sc Hyper-z}
code~\citep{Bolz00} and the same procedure described
in~\citet{Lons08}. In Fig. \ref{sed_fits_zphot_ol}, we show the optical-IR SED and the best-fit template and
corresponding photometric redshift of each source. In seven cases we also
report a secondary solution corresponding to a second minimum in the
$\chi^2$ distribution if associated with a different template than the
primary solution. In case a spectroscopic redshift is available (source
L-25), the best-fit template at the spectroscopic redshift is also reported.

\begin{figure*}

\includegraphics{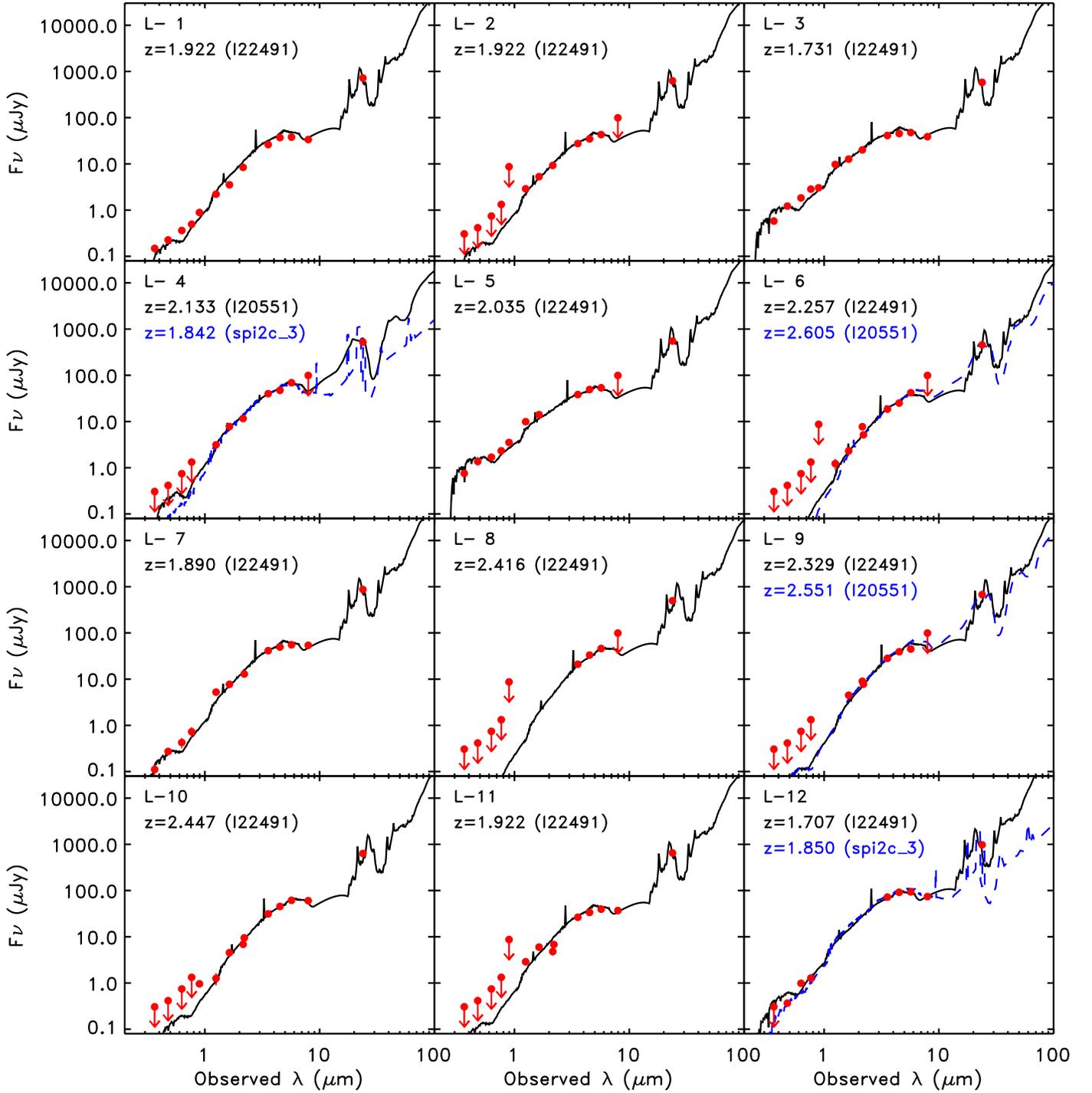}
 \caption{Optical-IR SEDs of our sources. The
 solid curves represent the best-fit template of the optical-MIR data (up to
 24\,$\mu$m). The dashed blue curves correspond to fits with increasing
 $\chi^2$.  The template name and photometric redshifts of each fit are
 annotated. For L-25, the dotted curve corresponds to the best-fit template plotted at
 the spectroscopic redshift. The source ID number 
  is reported in each
 panel. Downward arrows represent 5$\sigma$ upper limits at optical and
 infrared wavelengths.}
\label{sed_fits_zphot_ol}
\end{figure*}
\addtocounter{figure}{-1}
\begin{figure*}
\includegraphics{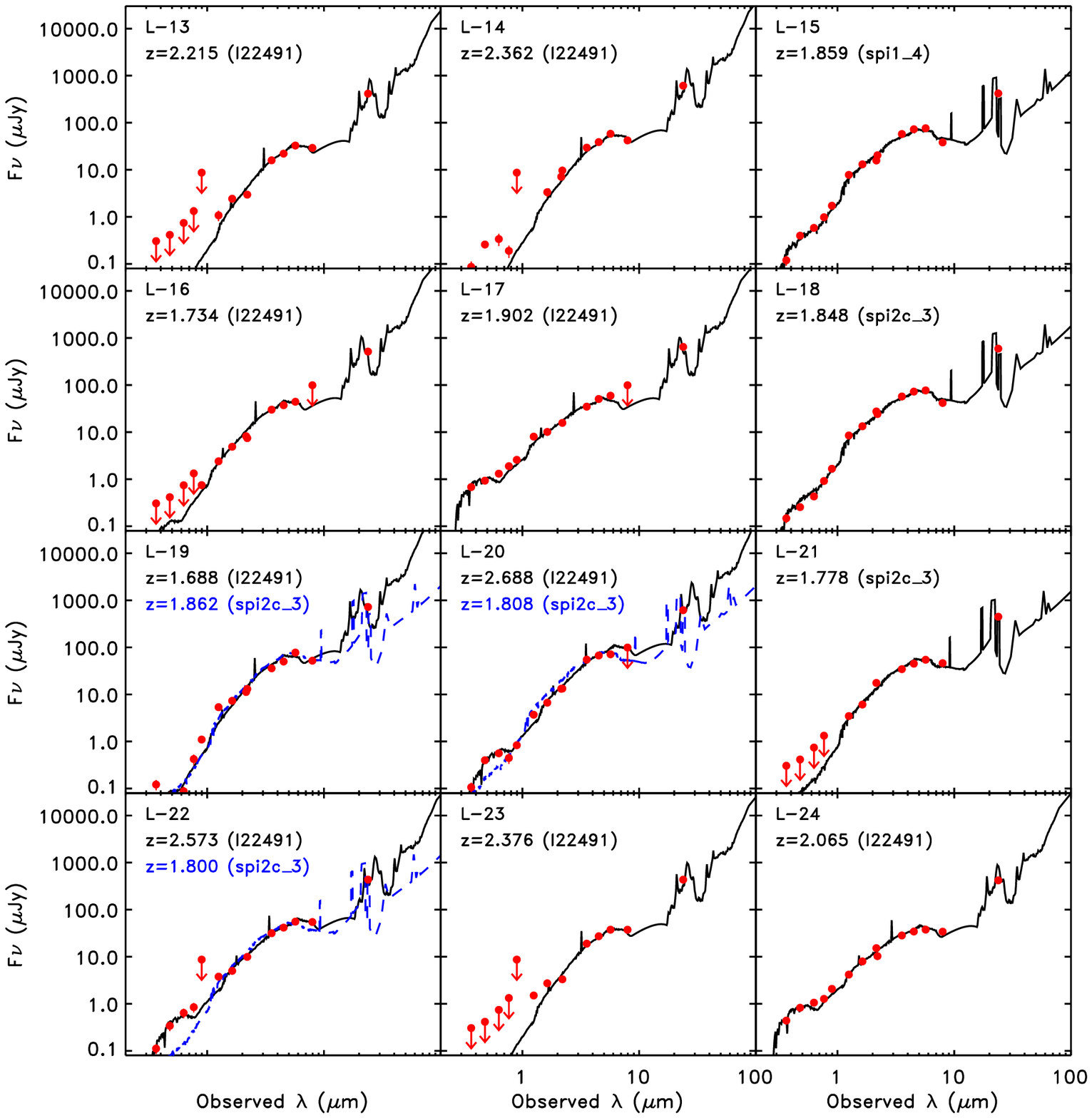}
 \caption{\it Continued}
\end{figure*}
\addtocounter{figure}{-1}
\begin{figure*}
\includegraphics{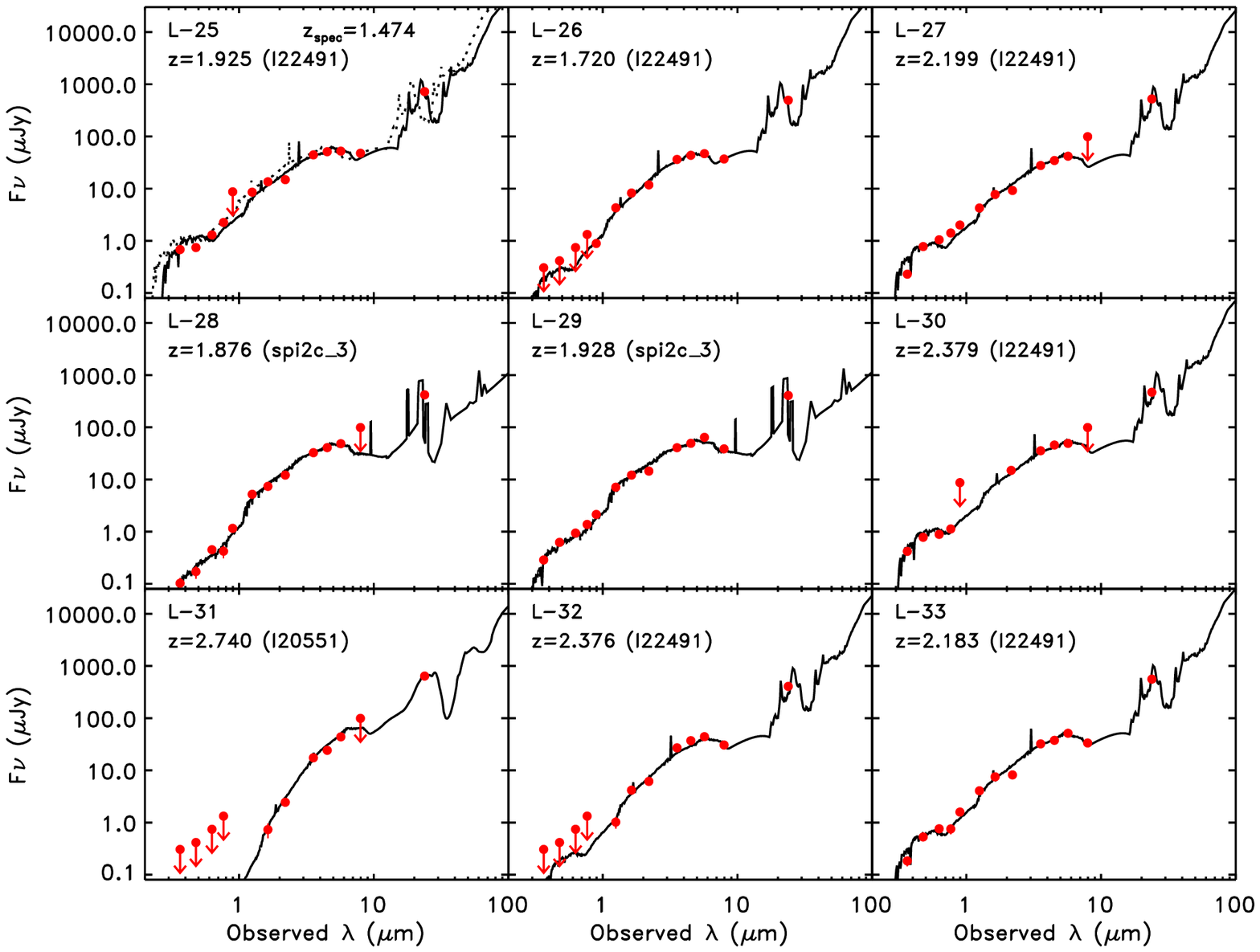}
 \caption{\it Continued}
\end{figure*}

\end{appendix}
\end{flushleft}  

\end{document}